\documentclass[a4paper]{jpconf}
\usepackage[utf8]{inputenc}
\usepackage{graphicx}
\usepackage{flafter}
\usepackage{amsmath}
\usepackage{amssymb}
\usepackage{float}
\usepackage{color}
\usepackage{subfigure}

\let \k \relax
\newcommand{\k}{{\bf k}}
\newcommand{\q}{{\bf q}}
\newcommand{\llangle}{\langle\langle}
\newcommand{\rrangle}{\rangle\rangle}
\newcommand{\pdagger}{{\phantom{\dagger}}}
\newcommand{\beps}{\bar\varepsilon}
\newcommand{\teps}{\tilde\varepsilon}
\let \Im \relax
\newcommand{\Im}{{\rm Im}}
\let \Re \relax
\newcommand{\Re}{{\rm Re}}

\begin{document}

\title{Exciton formation in strongly correlated electron-hole systems near the semimetal-semiconductor transition}
\author{B Zenker$^1$, D Ihle$^2$, F X Bronold$^1$ and H Fehske$^{1}$}
\address{$^1$Institut f{\"u}r Physik, Ernst-Moritz-Arndt-Universit{\"a}t Greifswald,
D-17489 Greifswald, Germany}
\address{$^2$Institut f{\"u}r Theoretische Physik, Universit{\"a}t Leipzig, 
D-04109 Leipzig, Germany}

\ead{zenker@physik.uni-greifswald.de, fehske@physik.uni-greifswald.de}

\date{\today}
\begin{abstract}
The region surrounding the excitonic insulator phase is a three-component plasma composed of electrons, holes, and excitons. Due to the extended nature of the excitons, their presence influences the surrounding electrons and holes. We analyze this correlation. To this end, we calculate the density of bound electrons, the density of electrons in the correlated state, the momentum-resolved exciton density, and the momentum-resolved density of electron-hole pairs that are correlated but unbound. We find qualitative differences in the electron-hole correlations between the weak-coupling and the strong-coupling regime.
\end{abstract}

\section{Introduction}
The semimetal-semiconductor (SM-SC) transition is of particular interest, since the excitonic insulator (EI) might be realized in its vicinity at low enough temperatures~\cite{Mo61, Kno63, KK65, JRK67, HR68}. Promising materials for the experimental verification are TmSe$_{0.45}$Te$_{0.55}$~\cite{BSW91}, 1$T$-TiSe$_2$~\cite{CMCBDGBAPBF07}, Ta$_2$NiSe$_5$~\cite{WSTMANTKNT09}, or a double bilayer graphene system~\cite{PNH13}.
 The EI constitutes an exciton condensate as an equilibrium phenomenon in contrast to optically created exciton condensates~\cite{LEKMSS04}. Approaching, respectively, the EI from the SM or the SC side, a BCS-type transition of electron-hole pairs or a Bose-Einstein condensation (BEC) of preformed excitons occurs. Hence, the EI is discussed in view of a BCS-BEC crossover~\cite{BF06, IPBBF08, ZIBF12}. A different nature of the electron-hole pairs below the critical temperature $T_{\rm EI}$ might also show up in terms of precursor effects above $T_{\rm EI}$~\cite{ZIBF12}. Close to $T_{\rm EI}$ free excitons appear possibly in a large number, and the region surrounding the EI, in Ref.~\cite{BF06} termed ``halo", constitutes a three-component plasma consisting of electrons, holes, and excitons. 
Whereas the EI transition scenario itself and the presence of the halo phase above $T_{\rm EI}$ was analyzed in previous works~\cite{BF06, IPBBF08, ZIBF12, ZIBF11, PFB11}, the correlation effects caused by the excitons have not been studied so far. We address this issue in this work.

The Falicov-Kimball model extended by a finite $f$-bandwidth, in the following shortly denoted as extended Falicov-Kimball model (EFKM), is the minimal model describing the SM-SC transition and the EI formation~\cite{IPBBF08,Ba02b,BGBL04,SC08}. Expressing the orbital degree of freedom by a pseudospin variable $\sigma$, the EFKM can be written as
\begin{equation}
H=\sum_{\k,\sigma} \varepsilon_{\k\sigma} c_{\k\sigma}^\dagger c_{\k\sigma}^\pdagger
+U \sum_i n_{i\uparrow} n_{i\downarrow},
\label{EFKM}
\end{equation}
where $c_{\k\sigma}^{(\dagger)}$ annihilates (creates) an electron with momentum $\k$ in the band labeled by $\sigma=\uparrow,\downarrow$, and $n_{i\sigma}=c_{i\sigma}^\dagger c_{i\sigma}^\pdagger$ characterizes the $\sigma$-electron occupation of the Wannier site $i$. Hereafter, $\sigma=\uparrow$ denotes the valence band and $\sigma=\downarrow$ denotes the conduction band. The bare band dispersions are given by $\varepsilon_{\k\sigma}$, the Coulomb interaction strength is denoted by $U$, and $N$ is the total number of lattice sites. In order to investigate the SM-SC transition we consider the EFKM at half filling,
\begin{equation}
n_\uparrow + n_\downarrow = 1,
\label{half_filling}
\end{equation}
where $n_\sigma=\frac{1}{N}\sum_\k \langle c_{\k\sigma}^\dagger c_{\k\sigma}^\pdagger \rangle$.

The scenario obtained within mean-field approximation agrees qualitatively with the result from more accurate approaches~\cite{ZIBF11, PFB11, ZIBF10, PBF10, SEO11}. The mean-field ground-state phase diagram is even in quantitative accordance with the constraint path Monte Carlo data~\cite{Fa08}. However, correlation effects are widely ignored by mean-field approaches, and for the detailed analysis of the bound states in the normal phase a more sophisticated approach is necessary. In this work we will apply the projection technique for Green functions developed by Plakida~\cite{Pl11} to describe the correlations in the ``electron-hole-exciton system" in terms of a self-energy.

The paper is organized as follows. In Sec.~\ref{sec:ProjectTech} we outline the method. In Sec.~\ref{sec:ehCorrs} we analyze the electron-hole correlations and determine the excitonic susceptibility, the density of bound electrons, and the density of electrons that are correlated but not bound. Our numerical results are presented in Sec.~\ref{sec:NumRes}.  Section~\ref{sec:Summary} summarizes this work.

\section{Projection technique for Green functions}
\label{sec:ProjectTech}
The projection approach starts with the decomposition of the time derivatives of the electron operators into a part proportional to themselves and an irreducible part (throughout the paper we set $\hbar=1$), 
\begin{equation}
i\frac{d}{dt} c_{\k\sigma}^\pdagger \equiv i\dot{c}_{\k\sigma}^\pdagger = \beps_{\k\sigma} c_{\k\sigma}^\pdagger + i\dot{c}_{\k\sigma}^{\rm (ir)} 
\hspace{0.5cm} {\rm with} \hspace{0.5cm}
\langle [ i\dot{c}_{\k\sigma}^{\rm (ir)}, c_{\k\sigma}^\dagger ]_+\rangle =0.
\label{time_derive}
\end{equation}
Introducing the spectral moments $M_{\k\sigma} = \langle [c_{\k\sigma}^\pdagger , c_{\k\sigma}^\dagger ]_+\rangle$ and $M_{\k\sigma}' = \langle [i\dot{c}_{\k\sigma}^\pdagger , c_{\k\sigma}^\dagger ]_+\rangle$, we have $M_{\k\sigma}=1$ and $M_{\k\sigma}'=\beps_{\k\sigma}=\varepsilon_{\k\sigma}+Un_{-\sigma}$.

To derive an exact expression for the self-energy, we consider the equation of motion for the Green function $G_{\k\sigma}(t-t')=\llangle c_{\k\sigma}^\pdagger(t) ; c_{\k\sigma}^\dagger (t')\rrangle$.  Differentiating $G_{\k\sigma}(t-t')$ successively with respect to time $t$ and $t'$, we obtain a system of equations which in the Fourier representation reads
\begin{eqnarray}
(\omega-\beps_{\k\sigma})\llangle c_{\k\sigma}^\pdagger ; c_{\k\sigma}^\dagger \rrangle_\omega
&=& 1+ \llangle i\dot{c}_{\k\sigma}^{\rm (ir)} ; c_{\k\sigma}^\dagger \rrangle_\omega ,
\label{EOM_G}
\\
(\omega-\beps_{\k\sigma}) \llangle i\dot{c}_{\k\sigma}^{\rm (ir)} ; c_{\k\sigma}^\dagger \rrangle_\omega &=& T_{\k\sigma}(\omega),
\label{EOM_Gprime}
\end{eqnarray}
with the scattering matrix
\begin{equation}
T_{\k\sigma} = \llangle i\dot{c}_{\k\sigma}^{\rm (ir)} ; -i \dot{c}_{\k\sigma}^{\dagger \;{\rm (ir)}} \rrangle_\omega ,
\label{scatMatrix}
\end{equation}
where $i\dot{c}_{\k\sigma}^{\rm (ir)} = \frac{U}{N} \sum_\q c_{\k+\q\sigma}^\pdagger \rho_{-\q-\sigma}^\pdagger - U n_{-\sigma}^\pdagger c_{\k\sigma}^\pdagger$ and $\rho_{\q \sigma}^\pdagger=\sum_\k c_{\k\sigma}^\dagger c_{\k+\q\sigma}^\pdagger$.
Introducing the zeroth-order Green function $G_{\k\sigma}^{(0)}(\omega)=(\omega-\beps_{\k\sigma})^{-1}$, we can rewrite Eqs.~\eqref{EOM_G} and \eqref{EOM_Gprime} as
\begin{equation}
G_{\k\sigma}^\pdagger (\omega) = G_{\k\sigma}^{(0)}(\omega)+ G_{\k\sigma}^{(0)}(\omega) T_{\k\sigma}^\pdagger (\omega) G_{\k\sigma}^{(0)}(\omega).
\label{Dyson_T}
\end{equation}
The self-energy is defined by the Dyson equation
\begin{equation}
G_{\k\sigma}^\pdagger(\omega) = G_{\k\sigma}^{(0)}(\omega) + G_{\k\sigma}^{(0)}(\omega) \Sigma_{\k\sigma}^\pdagger (\omega) G_{\k\sigma}^\pdagger (\omega).
\label{Dyson}
\end{equation}
From Eqs.~\eqref{Dyson_T} and \eqref{Dyson} we get the relation between the self-energy and the scattering matrix,
\begin{equation}
T_{\k\sigma}^\pdagger (\omega) = \Sigma_{\k\sigma}^\pdagger (\omega) 
+ \Sigma_{\k\sigma}^\pdagger (\omega) G_{\k\sigma}^{(0)}(\omega) T_{\k\sigma}^\pdagger (\omega).
\label{T_Sigma}
\end{equation}
This equation shows that the self-energy is the ``proper part" of the scattering matrix~\eqref{scatMatrix}, which has no parts connected by a single zeroth-order Green function, i.e.,
\begin{equation}
\Sigma_{\k\sigma}(\omega) = \llangle i\dot{c}_{\k\sigma}^{\rm (ir)} ;
-i \dot{c}_{\k\sigma}^{\dagger \;{\rm (ir)}} \rrangle_\omega^{(p)} .
\label{self_energy}
\end{equation}
The spectral representation of the self-energy is given by
\begin{equation}
\Sigma_{\k\sigma}(\omega) = \int_{-\infty}^{\infty} \frac{d\bar\omega}{2\pi}
\frac{\Gamma_{\k\sigma}(\bar\omega)}{\omega - \bar\omega} 
\label{spectral_SE}
\end{equation}
with
\begin{eqnarray}
\Gamma_{\k\sigma}(\bar\omega) &=& (e^{\beta\bar\omega}+1) \int_{-\infty}^{\infty} dt \,
e^{i\bar\omega t} \, \frac{U^2}{N} K_{\k\sigma}(t) ,
\label{Gamma} \\
K_{\k\sigma}(t) &=& \frac{1}{N} \sum_{\q,\q'} \langle \rho_{\q'-\sigma}(0)
c_{\k+\q'\sigma}^\dagger (0) c_{\k+\q \sigma}^\pdagger (t) \rho_{-\q-\sigma}(t)\rangle^{(p)}.
\label{K}
\end{eqnarray}

Then the spectral function of the electrons reads
\begin{equation}
A_{\k\sigma}(\omega) = -2 \Im G_{\k\sigma}(\omega) = -2 \frac{\Sigma_{\k\sigma}''(\omega)}
{\big[\omega-\beps_{\k\sigma}-\Sigma_{\k\sigma}'(\omega)\big]^2 
+ \big[\Sigma_{\k\sigma}''(\omega)\big]^2} \;,
\label{spectral_func}
\end{equation}
where we have introduced the shorthand notation $\Sigma_{\k\sigma}'(\omega)=\Re \Sigma_{\k\sigma}(\omega)$ and $\Sigma_{\k\sigma}''(\omega) = \Im \Sigma_{\k\sigma}(\omega)$.
The renormalized dispersion is given by
\begin{equation}
\teps_{\k\sigma} = \beps_{\k\sigma} + \Sigma_{k\sigma}'(\omega)\big|_{\omega=\teps_{\k\sigma}}.
\label{teps}
\end{equation}
Let us emphasize that the equations up to Eq.~\eqref{teps} are exact. However, approximations become necessary if $\Sigma_{\k\sigma}$ is calculated.

\section{Electron-hole correlations}
\label{sec:ehCorrs}
In Eq.~\eqref{K}, a three-particle correlation function enters the self-energy. 
To proceed, we perform a two-time decoupling, because the ``proper part" of the correlation function~\eqref{K} must be considered.
Following the Martin-Schwinger decoupling scheme~\cite{MS59}, we identify four contributions,
\begin{equation}
K_{\k\sigma}^\pdagger (t)= K_{\k\sigma}^{(1)}(t) + K_{\k\sigma}^{(2)}(t) 
+ K_{\k\sigma}^{(3)}(t)+ K_{\k\sigma}^{(4)}(t) ,
\label{Kk}
\end{equation}
with
\begin{eqnarray}
K_{\k\sigma}^{(1)}(t) &=& -\frac{2}{N}\sum_{\k',\q} 
\langle c_{\k'-\sigma}^\pdagger (0) c_{\k'-\sigma}^\dagger (t) \rangle
\langle c_{\k'-\q-\sigma}^\dagger (0) c_{\k'-\q-\sigma}^\pdagger (t) \rangle
\langle c_{\k+\q\sigma}^\dagger (0) c_{\k+\q\sigma}^\pdagger (t) \rangle ,
\label{K1}  \nonumber \\
\\
K_{\k\sigma}^{(2)}(t) &=& \frac{1}{N} \sum_{\k',\q,\q'}
\langle c_{\k'+\q'-\sigma}^\pdagger (0) c_{\k+\q'\sigma}^\dagger (0)
c_{\k+\q\sigma}^\pdagger (t) c_{\k'+\q-\sigma}^\dagger (t)\rangle
\langle c_{\k'-\sigma}^\dagger(0) c_{\k'-\sigma}^\pdagger (t) \rangle ,
\label{K2}  \nonumber \\
\\
K_{\k\sigma}^{(3)}(t) &=& \frac{1}{N} \sum_{\k',\q,\q'}
\langle c_{\q'-\sigma}^\dagger (0) c_{\q+\q'-\sigma}^\pdagger (0)
c_{\k'-\sigma}^\dagger (t) c_{\k'-\q-\sigma}^\pdagger (t)\rangle
\langle c_{\k+\q\sigma}^\dagger (0) c_{\k+\q\sigma}^\pdagger (t) \rangle ,
\label{K3} \nonumber \\
\\
K_{\k\sigma}^{(4)}(t) &=& \frac{1}{N} \sum_{\k',\q,\q'}
\langle c_{\k'-\q'-\sigma}^\dagger (0) c_{\k+\q'\sigma}^\dagger (0)
c_{\k+\q\sigma}^\pdagger (t) c_{\k'-\q-\sigma}^\pdagger (t)\rangle
\langle c_{\k'-\sigma}^\pdagger (0) c_{\k'-\sigma}^\dagger (t) \rangle .
\label{K4} \nonumber \\
\end{eqnarray}
Since we are concentrating on excitonic electron-hole fluctuations, described by the term $K_{\k\sigma}^{(2)}(t)$, we further decouple $K_{\k\sigma}^{(3)}(t)$ and $K_{\k\sigma}^{(4)}(t)$ into products of single-particle correlation functions. This leads to  $K_{\k\sigma}^{(3)}(t)= K_{\k\sigma}^{(4)}(t)=-\frac{1}{2} K_{\k\sigma}^{(1)}(t)$. Then, $K_{\k\sigma}^\pdagger (t)=K_{\k\sigma}^{(2)}(t)$, and we end up with
\begin{equation}
\Gamma_{\k\sigma}(\bar\omega) = -\frac{2U^2}{N\pi} \sum_{\k'} \int_{-\infty}^{\infty} d\omega'
\big[ f(\omega')+p(\omega'-\bar\omega)\big] 
\Im \chi_{\k-\k'}^{-\sigma,\sigma}(\bar\omega-\omega') \Im G_{\k'-\sigma}(\omega'),
\label{SF_SE}
\end{equation}
where $f(\omega)$ is the Fermi function, $p(\omega)$ is the Bose function, and 
\begin{equation}
\chi_{\q}^{-\sigma,\sigma} (\omega) = \frac{1}{N} \sum_{\k,\k'} 
\llangle c_{\k-\sigma}^\dagger  c_{\k+\q\sigma}^\pdagger ; 
c_{\k'+\q\sigma}^\dagger c_{\k'-\sigma}^\pdagger \rrangle_\omega 
\label{chi}
\end{equation}
is the excitonic susceptibility. We will concentrate on the presence of excitons and their effect on the unbound electrons by including only the according contribution in Eq.~\eqref{chi}.

\subsection{Excitonic susceptibility}
\begin{figure}[h]
\centering
\includegraphics[width=0.8\linewidth]{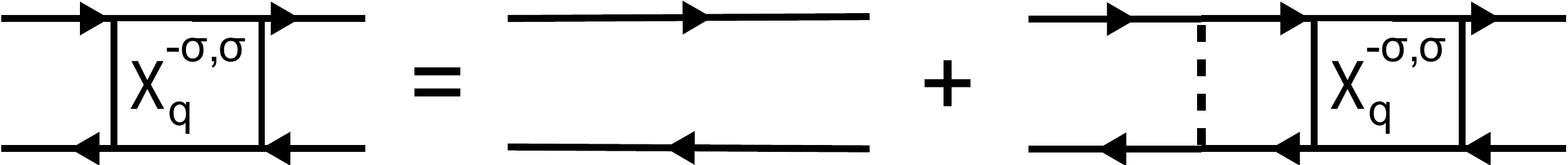}
\caption{Ladder approximation for the excitonic susceptibility $\chi_\q^{-\sigma,\sigma}$. The solid lines represent the electron Green functions, and the dashed line represents the Coulomb attraction.}
\label{fig:ladder}
\end{figure}

The ladder approximation~\cite{KB62} of the excitonic susceptibility, illustrated in Fig.~\ref{fig:ladder}, is suitable to describe electron-hole bound states.
Taking into account only the nearly-free part of the electron Green function, $G_{\k\sigma}^{\rm nf}(\omega) = (\omega-\teps_{\k\sigma})^{-1}$, we find
\begin{equation}
\chi_\q^{-\sigma,\sigma}(\omega) = \frac{\tilde\chi_\q^{-\sigma,\sigma}(\omega)}{U\tilde\chi_\q^{-\sigma,\sigma}(\omega) +1} ,
\label{eh_corr}
\end{equation}
where $\tilde\chi_\q^{-\sigma,\sigma}(\omega) =N^{-1} \sum_\k [f(\teps_{\k-\sigma})-f(\teps_{\k+\q\sigma})]/(\omega+\teps_{\k-\sigma}-\teps_{\k+\q\sigma})$.

The pole of the excitonic susceptibility located outside the electron-hole continuum,  $\omega_X^{-\sigma,\sigma}(\q)$ with $|\omega_X^{-\sigma,\sigma}(\q)|<|\omega_C^{-\sigma,\sigma}(\q)|= {\rm min}_\k |\teps_{\k+\q\sigma}-\teps_{\k-\sigma}|$, describes the bound state of an electron from the $\sigma$-band with a hole from the band with index $-\sigma$. 
Note that $\omega_X^{-\sigma,\sigma}(\q)$ increases with increasing temperature. However, for very high temperatures $\chi_\q^{-\sigma,\sigma}$ does not exhibit a pole outside the continuum: consequently excitons do not exist.

\subsection{Density of bound electrons}
Since an exciton contains one electron, the number of bound electrons equals the number of excitons.
Chosen $\sigma=\downarrow$ to be the index for the conduction band, the exciton dispersion is
\begin{equation}
\omega_X(\q) =\omega_X^{\uparrow,\downarrow}(\q)=-\omega_X^{\downarrow,\uparrow}(\q) .
\label{poles}
\end{equation} 
The binding energy of an exciton can be obtained from
$E_{\rm B}^X(\q) = \omega_C(\q)-\omega_X(\q)$,
where $\omega_C(\q)=\omega_C^{\uparrow,\downarrow}(\q)$.

Provided that $\omega_X(\q)$ exists, we follow Ref.~\cite{ZIBF12} and obtain the momentum-resolved exciton density  as
\begin{equation}
N_X(\q) =  Z(\omega_X,\q) p(\omega_X) ,
\label{Xnumbr}
\end{equation}
where the spectral weight of the excitons becomes
\begin{equation}
Z(\omega_X,\q) = \bigg[ \frac{U^2}{N}\sum_\k \frac{f(\teps_{\k\uparrow})-f(\teps_{\k+\q\downarrow})}{(\omega_X+\teps_{\k\uparrow}-\teps_{\k+\q\downarrow})^2} \bigg]^{-1}
=Z^{\uparrow,\downarrow}(\omega_X^{\uparrow,\downarrow},\q) 
=Z^{\downarrow,\uparrow}(\omega_X^{\downarrow,\uparrow},\q) ,
\label{Sp_weight}
\end{equation}
and the exciton density (density of bound electrons) is given by 
\begin{equation}
n_X=\frac{1}{N}\sum_\q N_X(\q).
\label{Xdensity}
\end{equation}

\subsection{Density of correlated electrons}
Expanding the spectral function, Eq.~\eqref{spectral_func}, for small damping~\cite{KKER86},  we can decompose the density into a nearly-free part (with renormalized band dispersion) and a part, where the exciton formation enters,
\begin{equation}
n_\sigma = \frac{1}{N}\sum_{\k} \int_{-\infty}^\infty \frac{d\omega}{2\pi} A_{\k\sigma}(\omega) f(\omega)
= n_\sigma^{\rm nf} + n_{\sigma}^{\rm corr},
\label{dens_split} 
\end{equation}
where $n_\sigma^{\rm nf}=\frac{1}{N} \sum_\k f(\teps_{\k\sigma})$, and the density of the correlated electrons reads
\begin{equation}
n_\sigma^{\rm corr} = {\rm sgn}( \omega_X^{-\sigma,\sigma}) \frac{U^2}{N^2}\sum_{\k,\k'}
E^{-\sigma,\sigma}(\k,\k') F^{-\sigma,\sigma}(\k,\k') .
\label{n_corr}
\end{equation}
Straightforward calculation shows that
\begin{eqnarray}
E^{\uparrow,\downarrow}(\k,\k') &=& \frac{Z(\omega_X,\k-\k')}
{(\omega_X+\beps_{\k'\uparrow}-\beps_{\k\downarrow})^2}
=E^{\downarrow,\uparrow}(\k',\k) ,
\label{Esigma}
\\
F^{\uparrow,\downarrow}(\k,\k') &=& f(\teps_{\k'\uparrow}) f(\teps_{\k\downarrow}) 
-f(\teps_{\k\downarrow}) - f(\teps_{\k\downarrow})p(\omega_X^{\uparrow,\downarrow})
+f(\teps_{\k'\uparrow}) p(\omega_X^{\uparrow,\downarrow})
=F^{\downarrow,\uparrow}(\k',\k) .
\label{Fsigma}
\end{eqnarray}
Now, it is easy to see that $n_{\uparrow}^{\rm corr}=-n_{\downarrow}^{\rm corr}$, and
the chemical potential is exclusively determined by the nearly-free part of the densities (see also~\cite{BF06}): 
\begin{equation}
n_\uparrow + n_\downarrow = n_{\uparrow}^{\rm nf} + n_{\downarrow}^{\rm nf} = 1.
\label{mu_det}
\end{equation}
Note that $n_\downarrow^{\rm corr} \neq n_X$, i.e., the considered correlation effects go beyond the simple binding of electrons (see below). Moreover, the  presence of excitons leads to a decrease of the density of valence-band electrons, because $n_\uparrow^{\rm corr}<0$.

\section{Numerical results}
\label{sec:NumRes}
Let us first recapitulate the basic properties of the EFKM. Thereby we restrict ourselves to a square lattice and a direct band-gap situation, i.e., valence-band maximum and conduction-band minimum lie at the Brillouin-zone center. We choose $t_\downarrow$ to be the unit of energy,  and the other model parameters are $E_\downarrow=0$, $E_\uparrow=-2.4$, and $t_\uparrow=-0.8$. In a first approximation we neglect the exciton-induced band renormalization in the numerics, i.e., $\teps_{\k\sigma}=\beps_{\k\sigma}$, where the mean-field Hartree-shift is contained in $\beps_{\k\sigma}$.
\begin{figure}[t]
\centering
\includegraphics[height=0.22\linewidth]{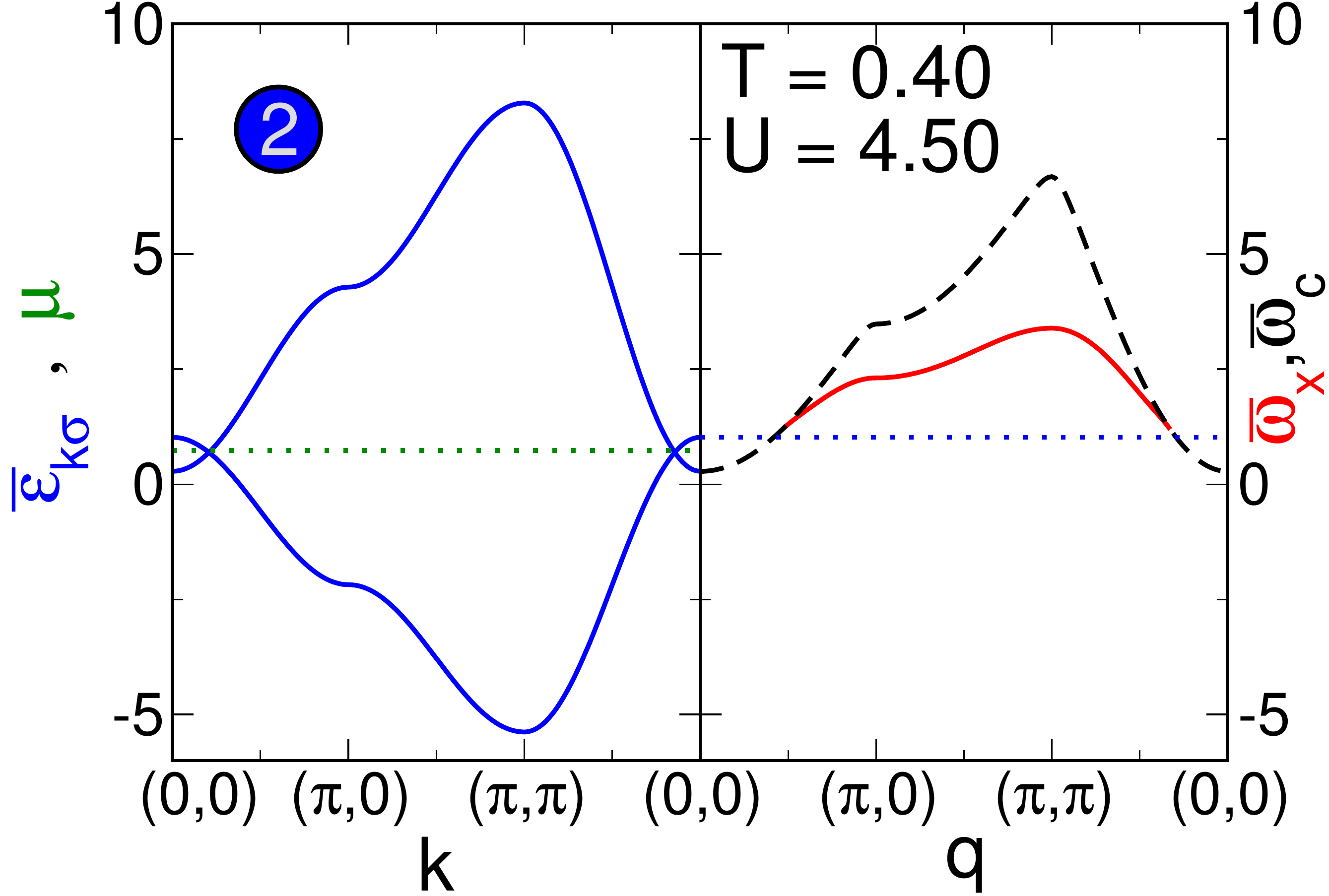}
\includegraphics[height=0.22\linewidth]{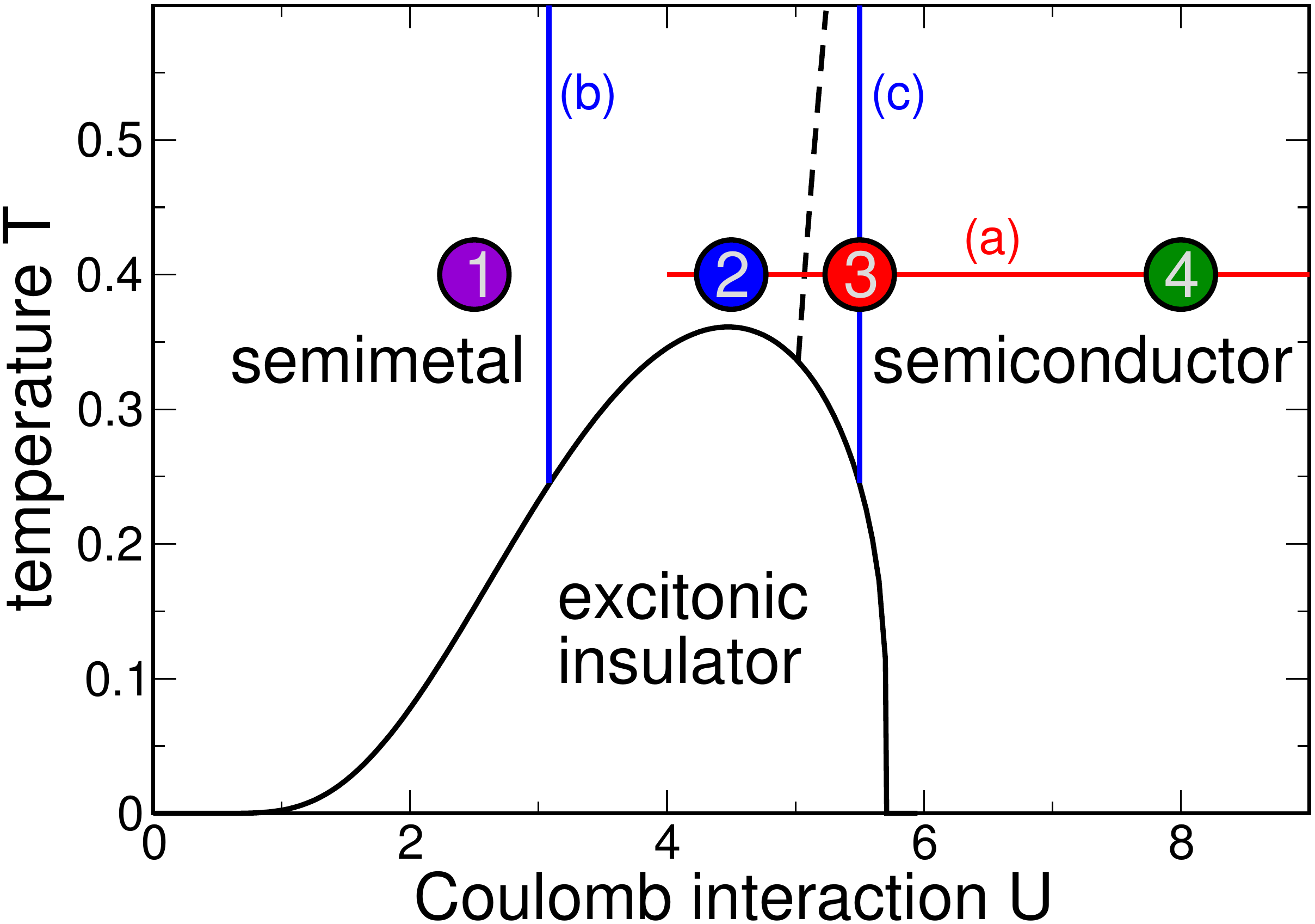}
\includegraphics[height=0.22\linewidth]{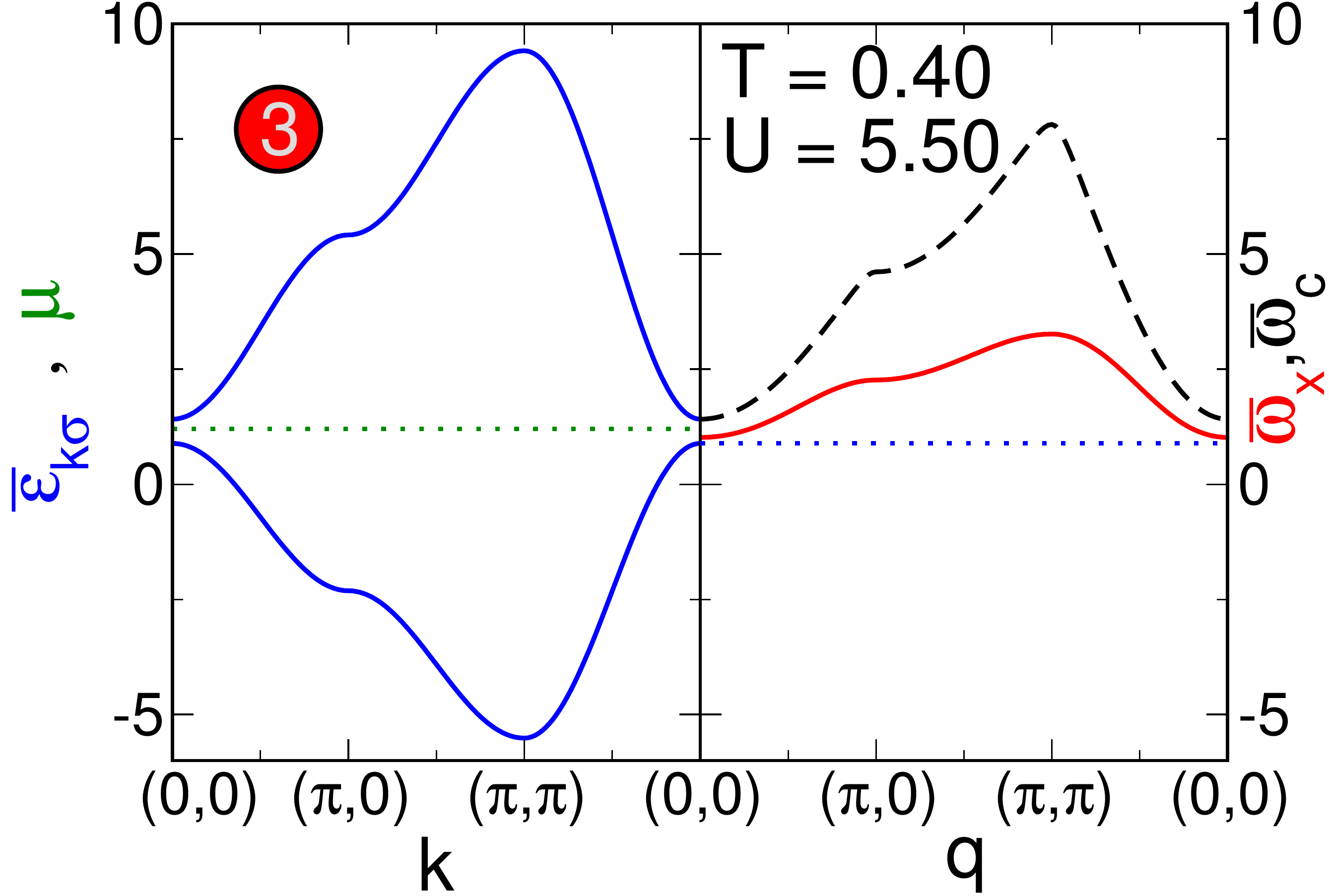}
\caption{(Middle panel) Phase diagram of the EI phase in the half-filled EFKM for a square lattice. In Fig.~\ref{nX_nCorr} we analyze the density of bound electrons and the density of correlated electrons along the red and blue lines (a), (b), and (c). The points 1, 2, 3, and 4 mark the points where we analyze the momentum-resolved exciton density and the momentum-resolved density of correlated but unbound electron-hole pairs in Fig.~\ref{X_corr}. (Left panel) Band structure and exciton dispersion for a semimetal. (Right panel) Band structure and exciton dispersion for a semiconductor. In the outer panels the renormalized band dispersions $\beps_{\k\sigma}$, the chemical potential $\mu$, the exciton energies $\bar\omega_X=\omega_X+{\rm max}_\k(\beps_{\k\uparrow})$, and the boundary of the electron-hole continuum $\bar\omega_C=\omega_C+{\rm max}_\k(\beps_{\k\uparrow})$ are shown. The blue dotted line shows ${\rm max}_\k(\beps_{\k\uparrow})$. The binding energy of an exciton is given by $E_B^X(\q)=\bar\omega_C(\q)-\bar\omega_X(\q)$. }
\label{PD_bandstructure}
\end{figure}

The middle panel of Fig.~\ref{PD_bandstructure} shows the phase diagram of the EFKM in the $U$-$T$ plane. The EI separates the SM phase and the SC phase at low temperatures. Valence band and conduction band overlap in a SM and, as a result, a well-defined Fermi surface exists. In this case, only finite momentum excitons can develop, see Fig.~\ref{PD_bandstructure} (left panel). On the other hand, in a SC the valence band and the conduction band are separated by an energy gap, and excitons with an arbitrary momentum can form, in particular zero-momentum excitons, which undergo a BEC at the SC-EI transition, see Fig.~\ref{PD_bandstructure} (right panel).

Electrons and holes may bind to excitons, and subsequently the remaining unbound electrons and holes may scatter as well on these quasiparticles. Both effects, the exciton binding and the influence on the  unbound electrons, are considered in $n_\sigma^{\rm corr}$. How strongly the electrons are affected is mainly determined by the spatial extension of the excitons, and the underlying band structure. 

\begin{figure}[t]
\centering
\subfigure{
\includegraphics[width=0.31\linewidth]{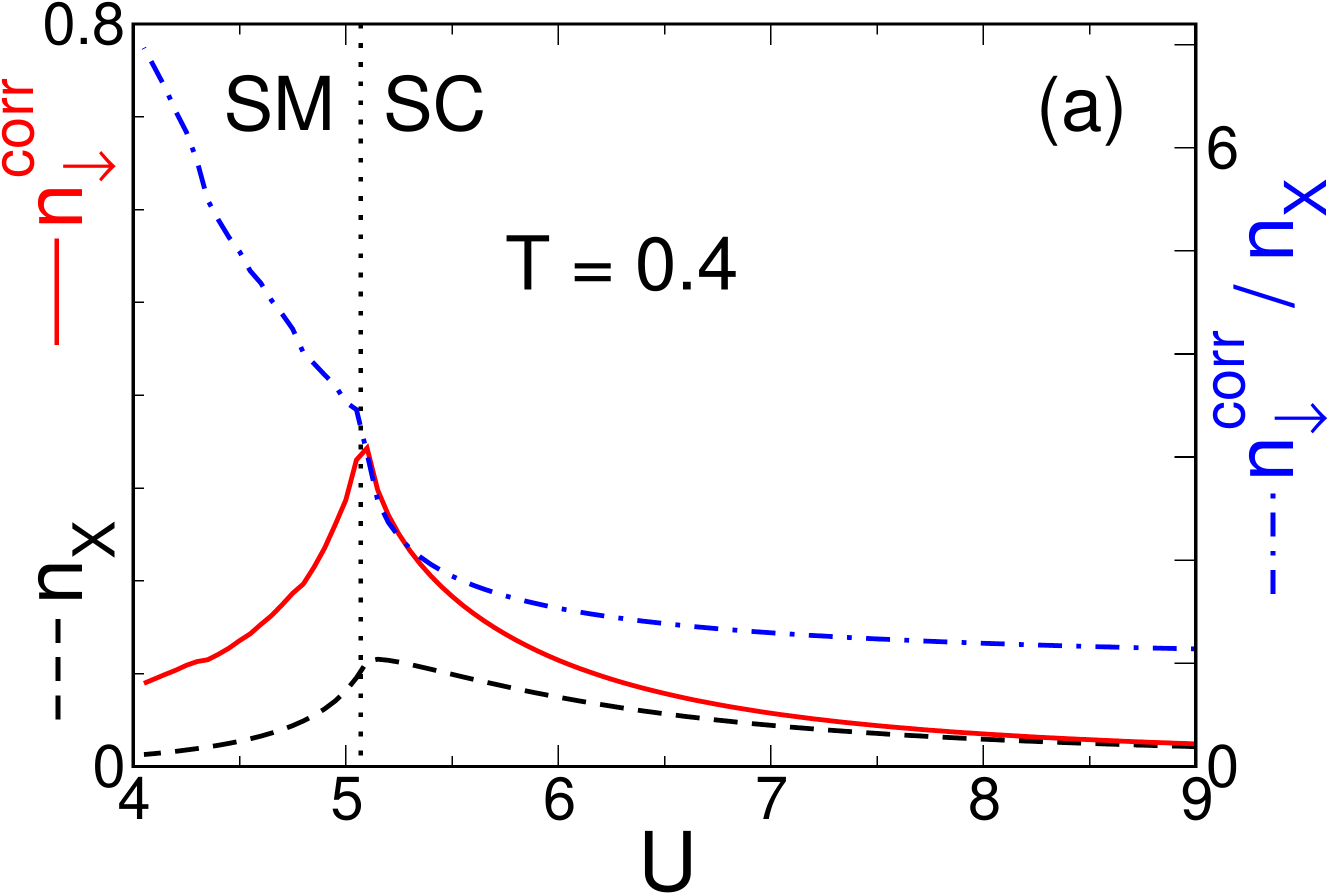}
}
\subfigure{
\includegraphics[width=0.31\linewidth]{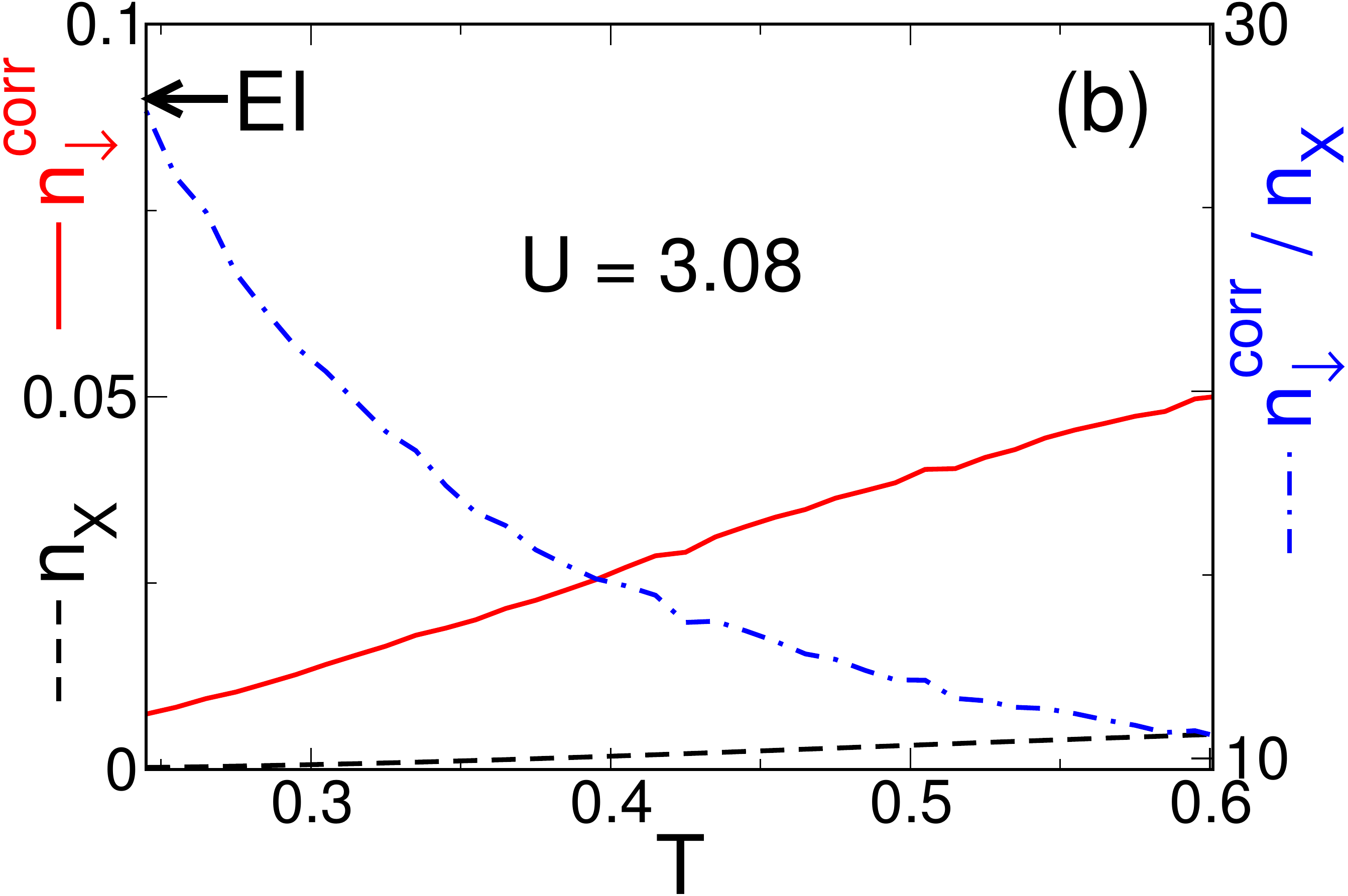}
}
\subfigure{
\includegraphics[width=0.31\linewidth]{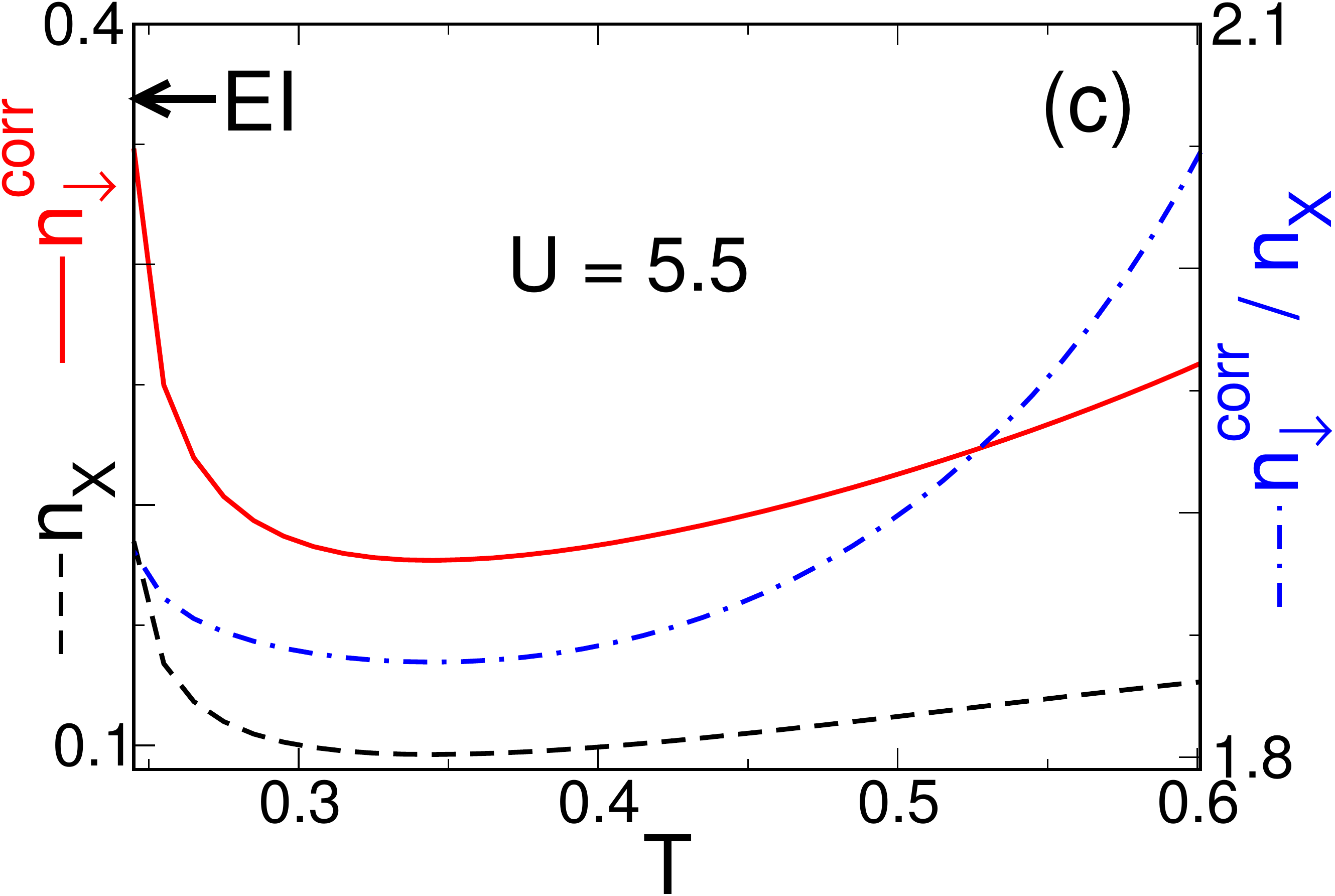}
}
\caption{Density of bound electrons $n_X$ (black, dashed line), density of correlated electrons $n_\downarrow^{\rm corr}$ (red, solid line), and the ratio between them $n_X/n_\downarrow^{\rm corr}$ (blue, dot-dashed line) (a) as a function of the Coulomb interaction $U$ ($T=0.4$), crossing thereby the semimetal-semiconductor transition , (b) as a function of temperature for a semimetal ($U=3.08$), and (c) as a function of temperature for a semiconductor ($U=5.5$). The panels (a), (b), (c) correspond to the lines indicated in the middle panel of Fig.~\ref{PD_bandstructure}.}
\label{nX_nCorr}
\end{figure}

Let us start by keeping the temperature constant and raising $U$, crossing thereby the SM-SC transition, see Fig.~\ref{nX_nCorr}(a). On the SM side, excitons are loosely bound objects with a relatively large extension. As a result, many electrons are influenced by a single exciton.
An increasing Coulomb attraction $U$ causes a tighter binding of electrons and holes. Consequently the exciton radius shrinks, and $n_\downarrow^{\rm corr}/n_X$ decreases.

At the SM-SC transition both the density of bound electrons and the density of correlated electrons reach their maximum. The kink of $n_X$ and $n_\downarrow^{\rm corr}$ will be less pronounced for a three-dimensional system, c.f. Ref.~\cite{ZIBF12}.

In the strong coupling regime, $n_\downarrow^{\rm corr}$ tends to $n_X$.
Here, the large Coulomb attraction gives rise to a very small distance between the bound electrons and holes: the Frenkel-limit of excitons is realized. 
Since there is only a very limited number of unbound electrons and holes, the correlation effects---apart from those leading to the formation of excitons by itself---are rather weak.

Let us now analyze temperature effects. On the SM side, excitons are less important for the EI formation~\cite{ZIBF12} and only a few electrons are bound or correlated, see Fig.~\ref{nX_nCorr}(b). With increasing temperature the broadening of the Fermi and Bose function will provide more electrons and holes to form excitons, which overcompensates the exciton dissociation. 
However, for very high temperatures the situation reverses, and the exciton dissociation dominates: $n_X$ and $n_{\downarrow}^{\rm corr}$ decrease with growing temperature until excitons disappear completely at $T_X=6.7$ for $U=3.08$ [not shown in Fig.~\ref{nX_nCorr}(b)].  
The higher kinetic energy at high temperatures enables the electrons to avoid correlations, and the ratio $n_\downarrow^{\rm corr} / n_X$ decreases with increasing temperature.

On the SC side, other effects prevail, see Fig.~\ref{nX_nCorr}(c). The SC-EI transition at $T_{\rm EI}$ is driven by the BEC of zero-momentum excitons. Close to $T_{\rm EI}$ the proliferation of $q=0$ excitons leads to a large density of electrons that are bound or correlated. 
In contrast to the SM side, the ratio $n_\downarrow^{\rm corr} / n_X$ increases with increasing temperature.
Here, the shrinking of the band gap makes conduction-band electrons available which can participate in the correlations. Note that $n_\downarrow^{\rm corr} / n_X$ is now significantly smaller than on the SM side, suggesting that the exciton formation resembles more the creation of local electron-hole pairs.
For $U=5.5$ [Coulomb strength in Fig.~\ref{nX_nCorr}(c)] the SC-SM transition is crossed at $T_{\rm SC-SM}=0.74$, and the total density of excitons vanishes at $T_X=12.5$.

Although the constraint for the chemical potential reduces to the mean-field condition, the electron densities differ significantly from their mean-field values, see Eq.~\eqref{dens_split}. 

Having examined the composition of the halo phase, we turn to the momentum distribution of  bound and unbound electron-hole pairs. By the replacement $\k'\rightarrow \k+\q$ in Eq.~\eqref{n_corr} we can write $n_\downarrow^{\rm corr}=N^{-1}\sum_\q N_\downarrow^{\rm corr}(\q)$, with
\begin{equation}
N_\downarrow^{\rm corr}(\q) = \frac{U^2}{N} \sum_\k Z(\omega_X,\q) 
\frac{ f(\teps_{\k+\q\downarrow}) [f(\teps_{\k\uparrow})-1]}
{(\omega_X + \teps_{\k\uparrow}-\teps_{\k+\q\downarrow})^2}
+ p(\omega_X) .
\label{Corrnumbr}
\end{equation}
We separate $N_\downarrow^{\rm corr}(\q)$ into bound pairs (excitons) and pairs which contain conduction-band electrons and valence-band holes that are correlated but unbound. We will shortly denote these objects as correlated pairs. The momentum-resolved exciton density $N_X(\q)$ is given by Eq.~\eqref{Xnumbr}, and the momentum-resolved density of correlated pairs is given by
\begin{equation}
D_X(\q) = N_\downarrow^{\rm corr}(\q) - N_X(\q) . 
\label{DCX}
\end{equation}
\begin{figure}[t]
\centering
\begin{minipage}{0.05\linewidth}
\includegraphics[width=\linewidth]{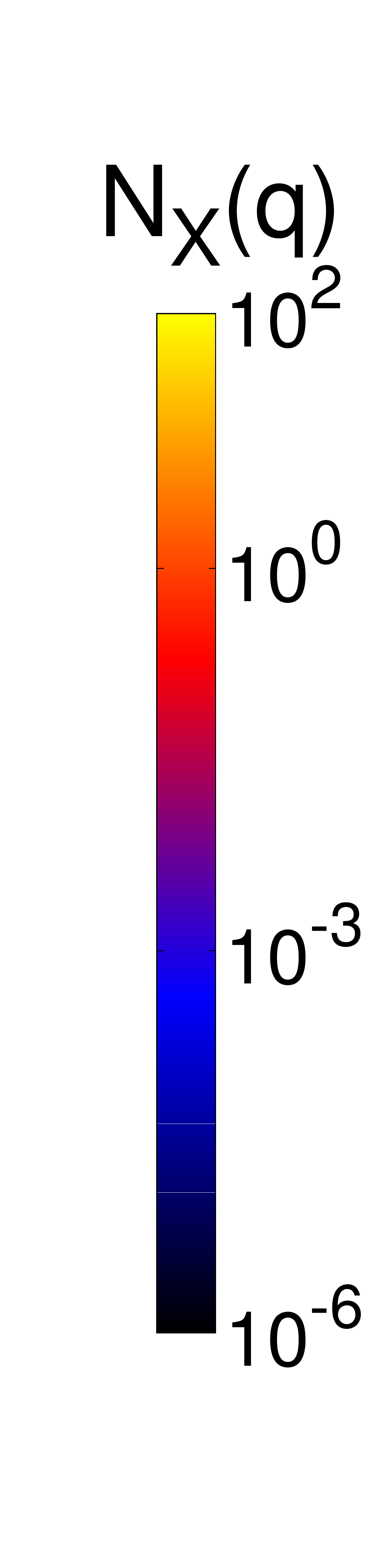} \\
\includegraphics[width=\linewidth]{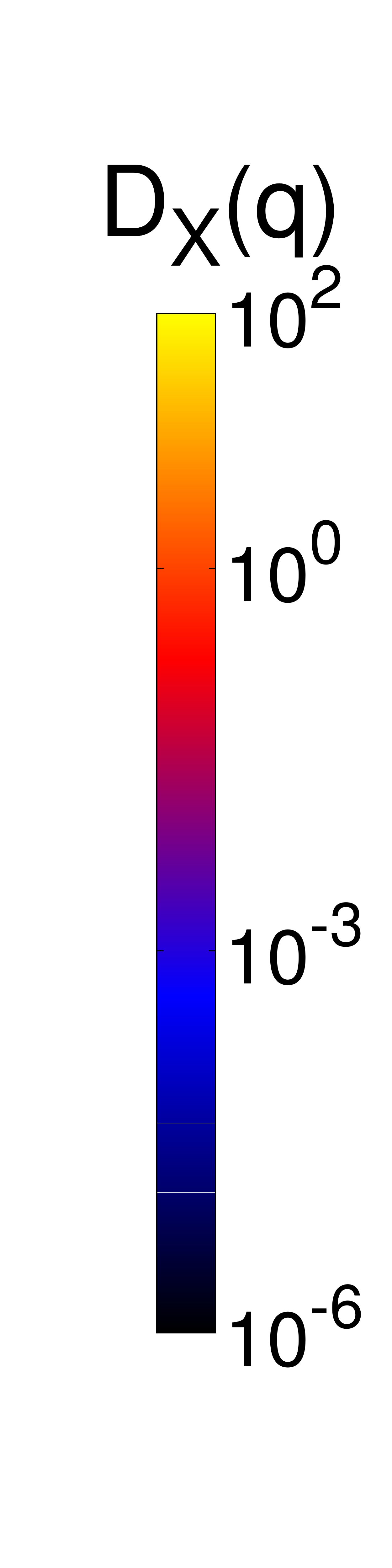} 
\end{minipage}
\begin{minipage}{0.21\linewidth}
\includegraphics[width=\linewidth]{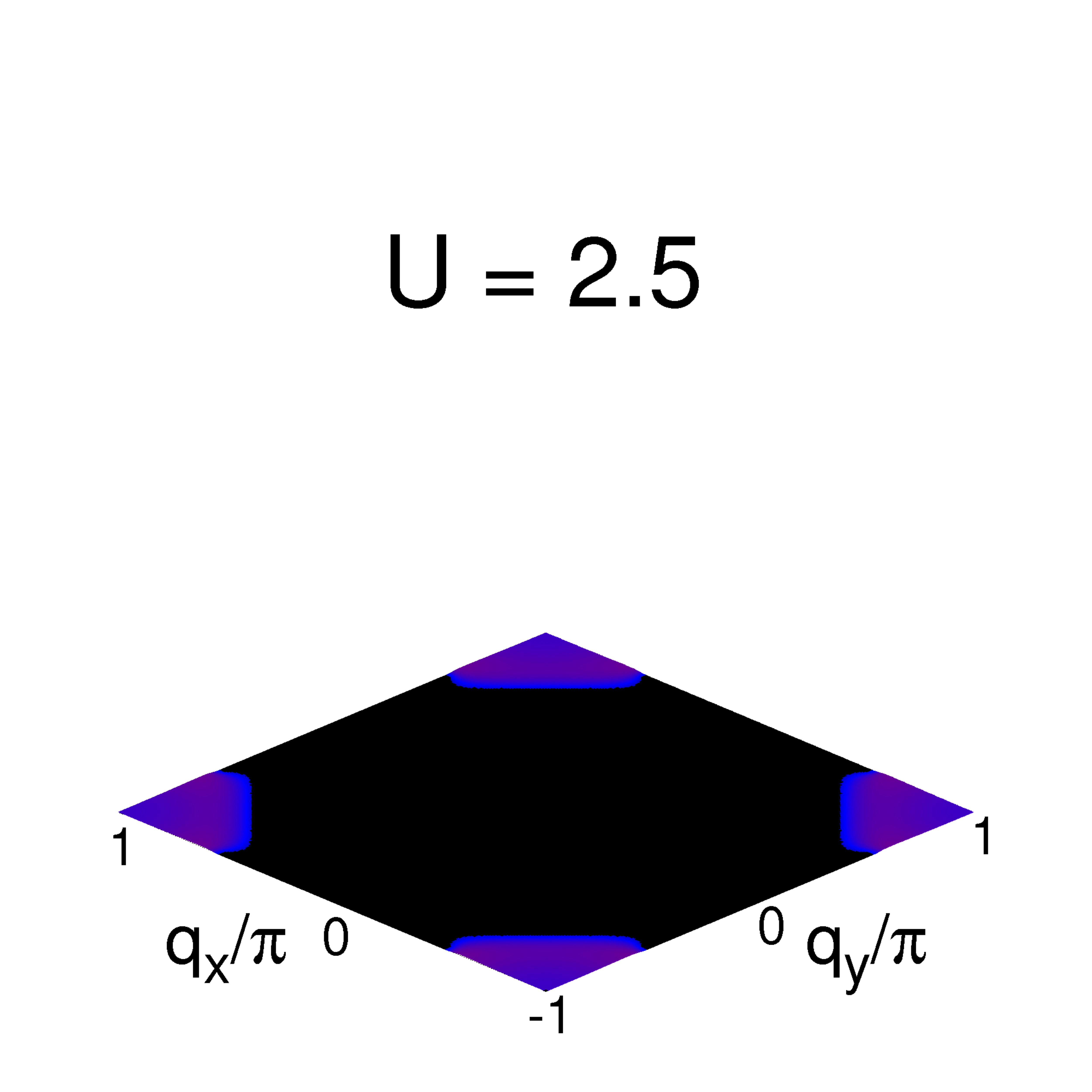} \\
\includegraphics[width=\linewidth]{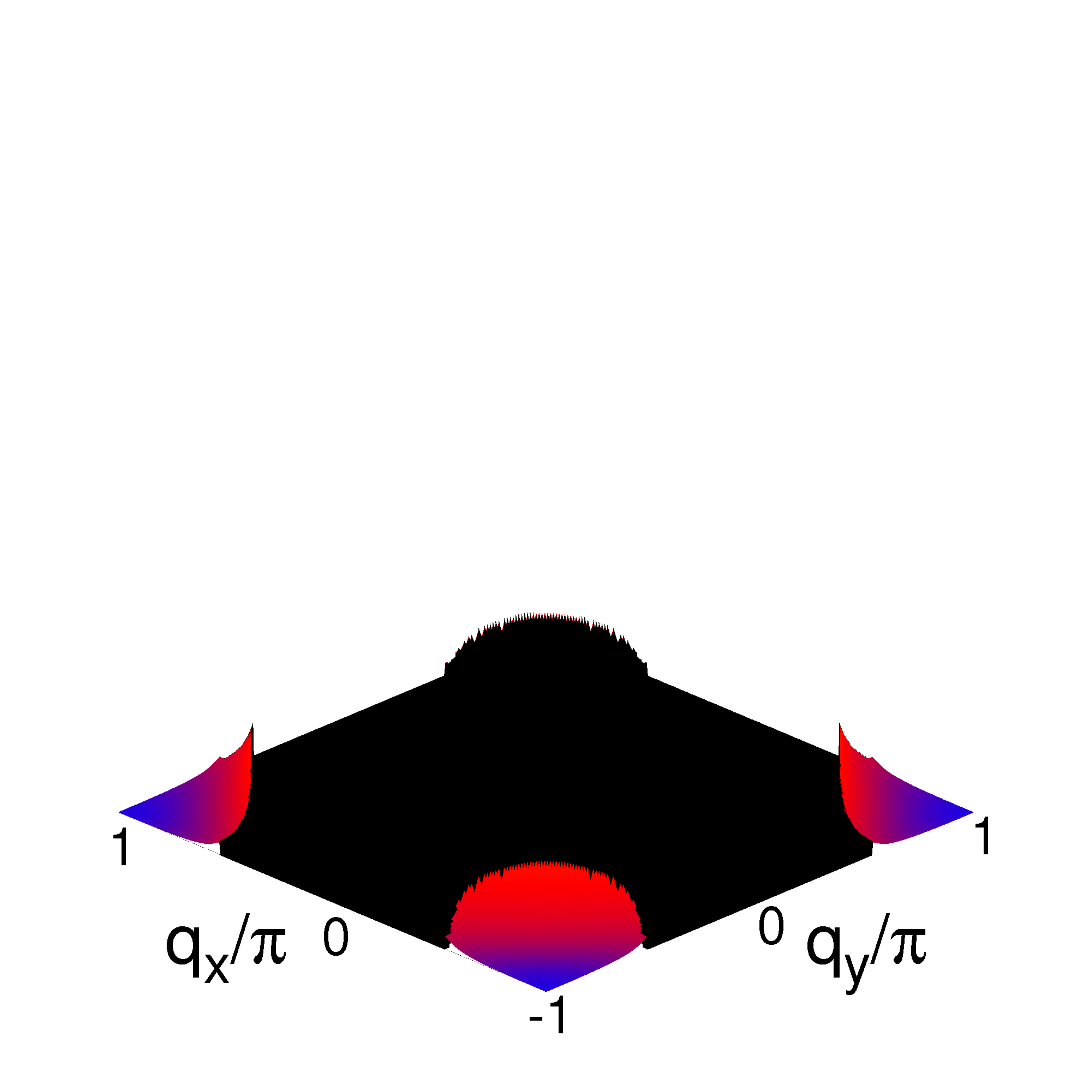} 
\end{minipage}
\begin{minipage}{0.21\linewidth}
\includegraphics[width=\linewidth]{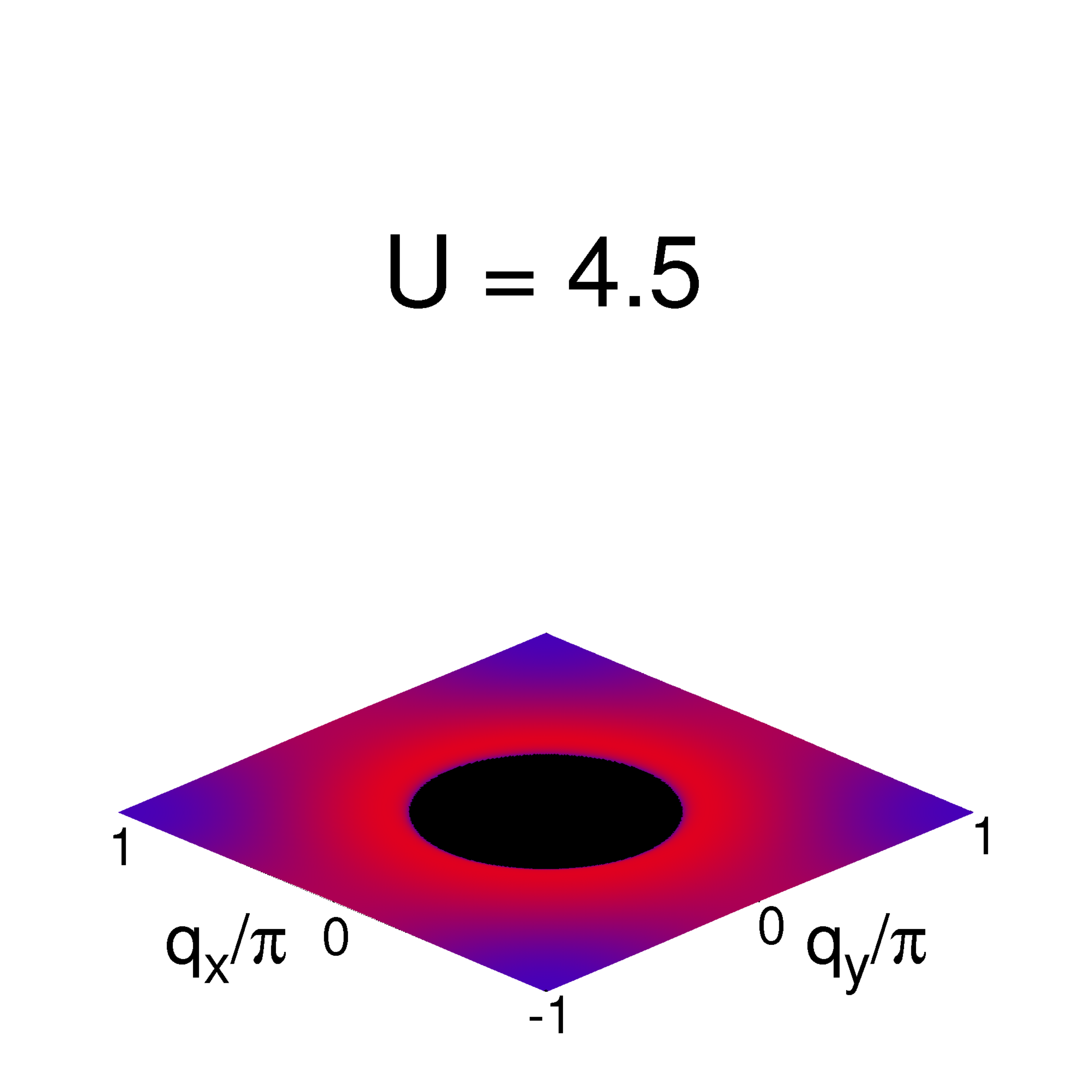} \\
\includegraphics[width=\linewidth]{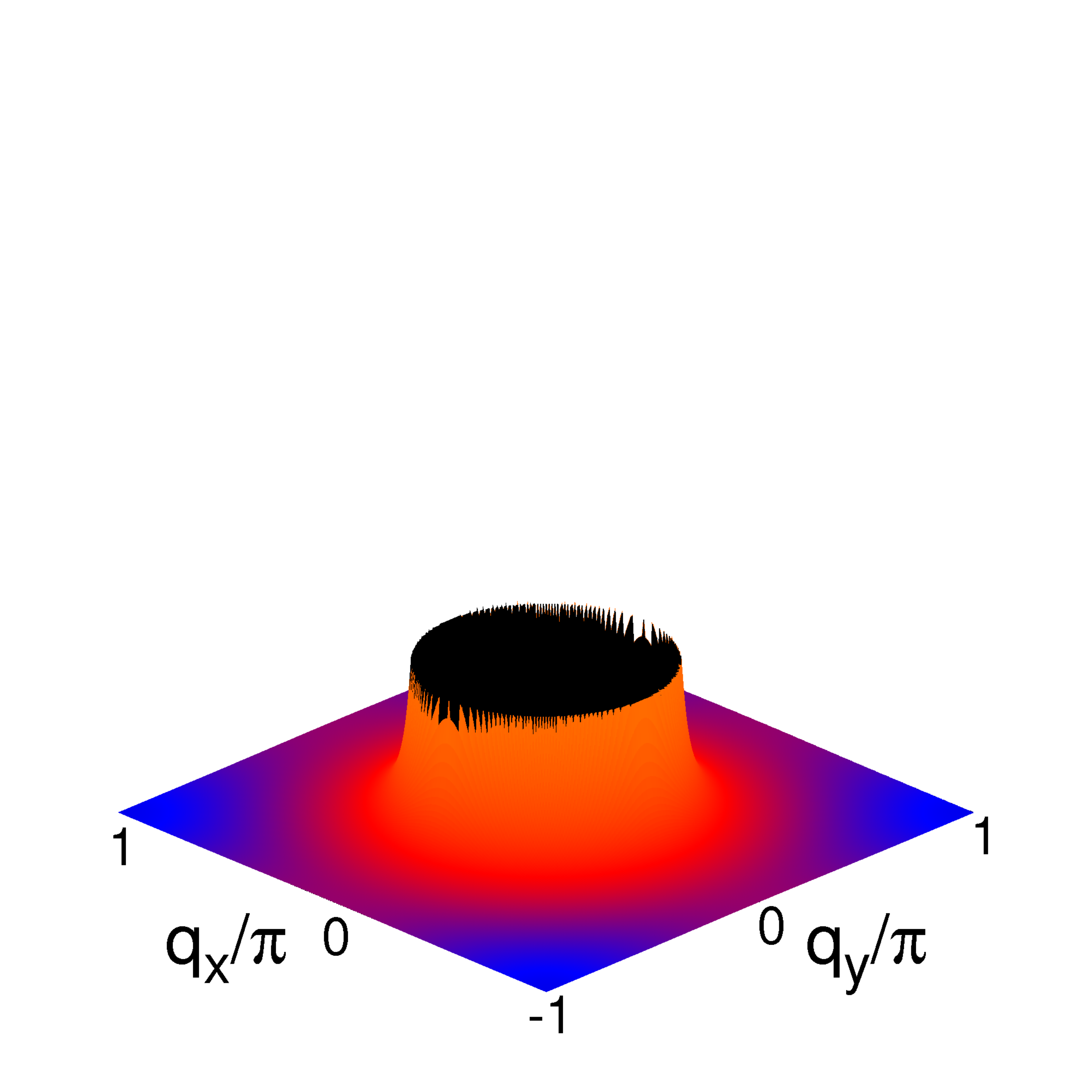} 
\end{minipage}
\begin{minipage}{0.21\linewidth}
\includegraphics[width=\linewidth]{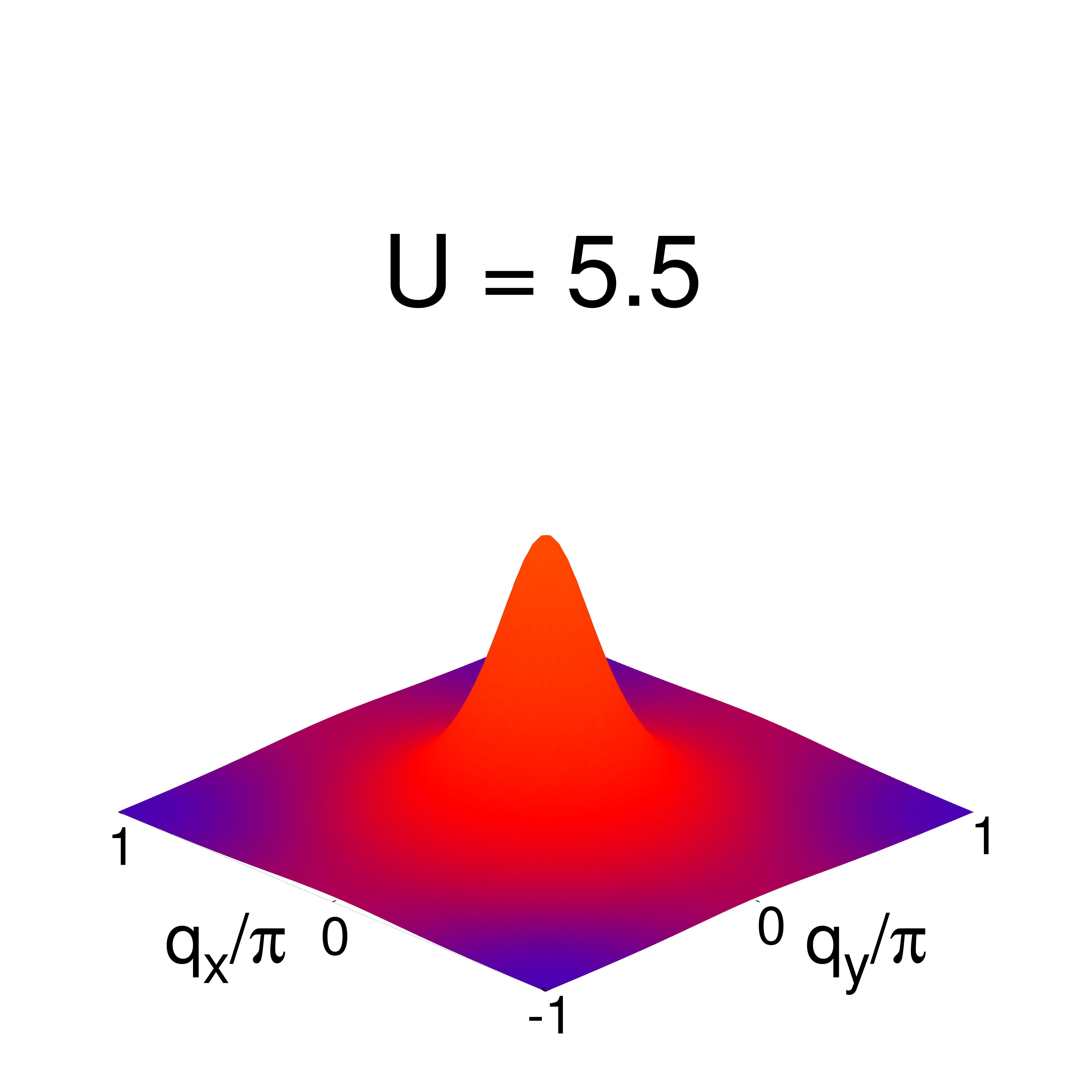} \\
\includegraphics[width=\linewidth]{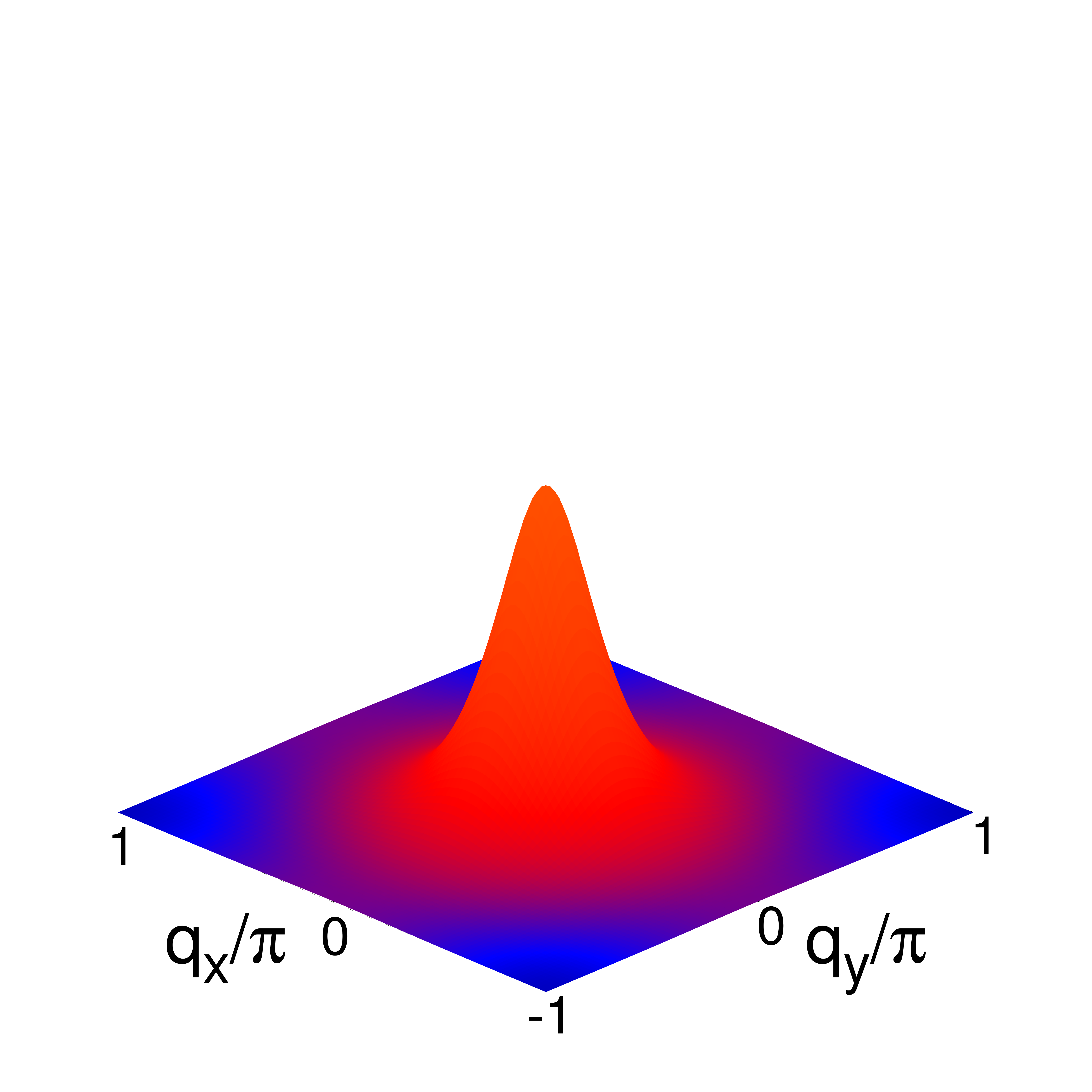} 
\end{minipage}
\begin{minipage}{0.21\linewidth}
\includegraphics[width=\linewidth]{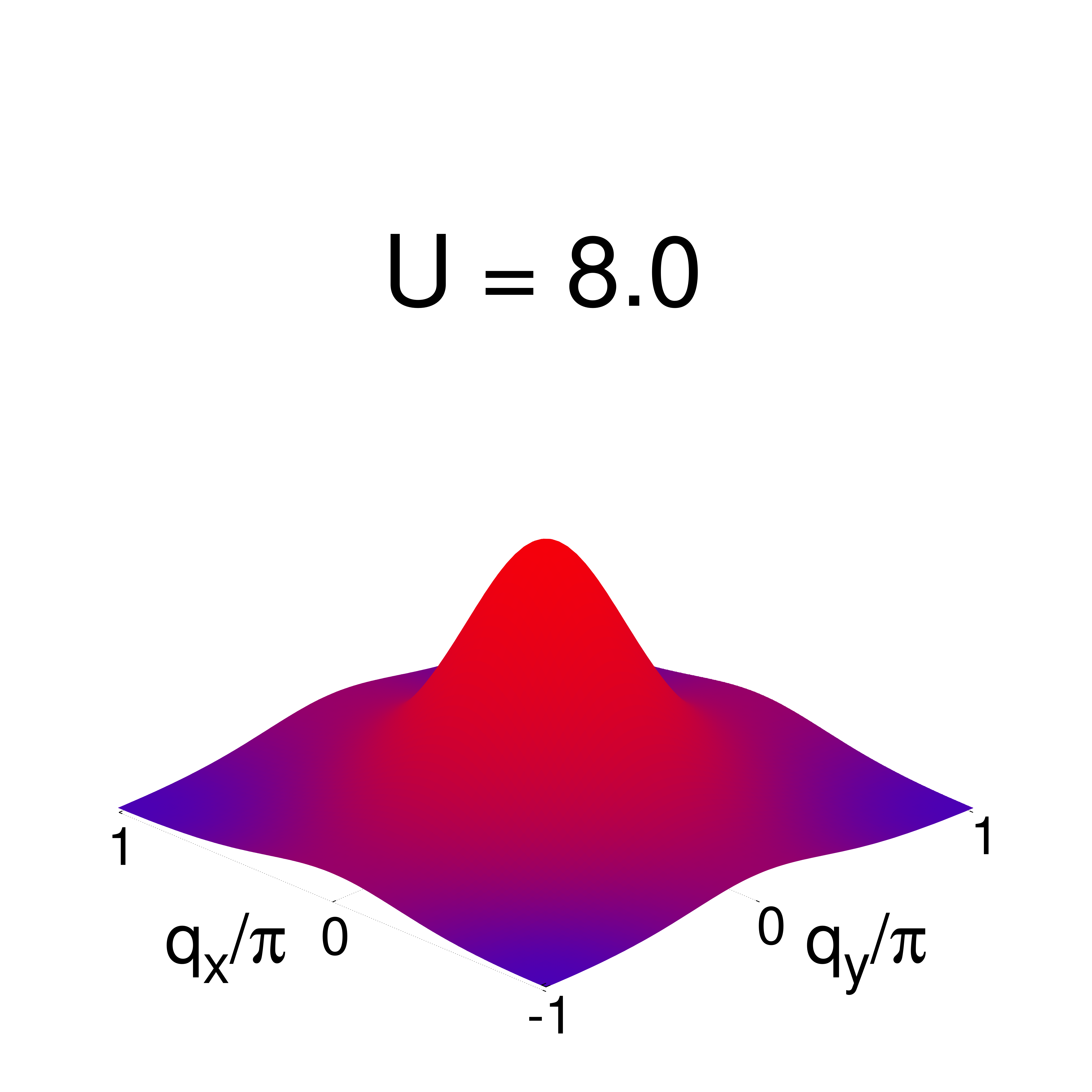} \\
\includegraphics[width=\linewidth]{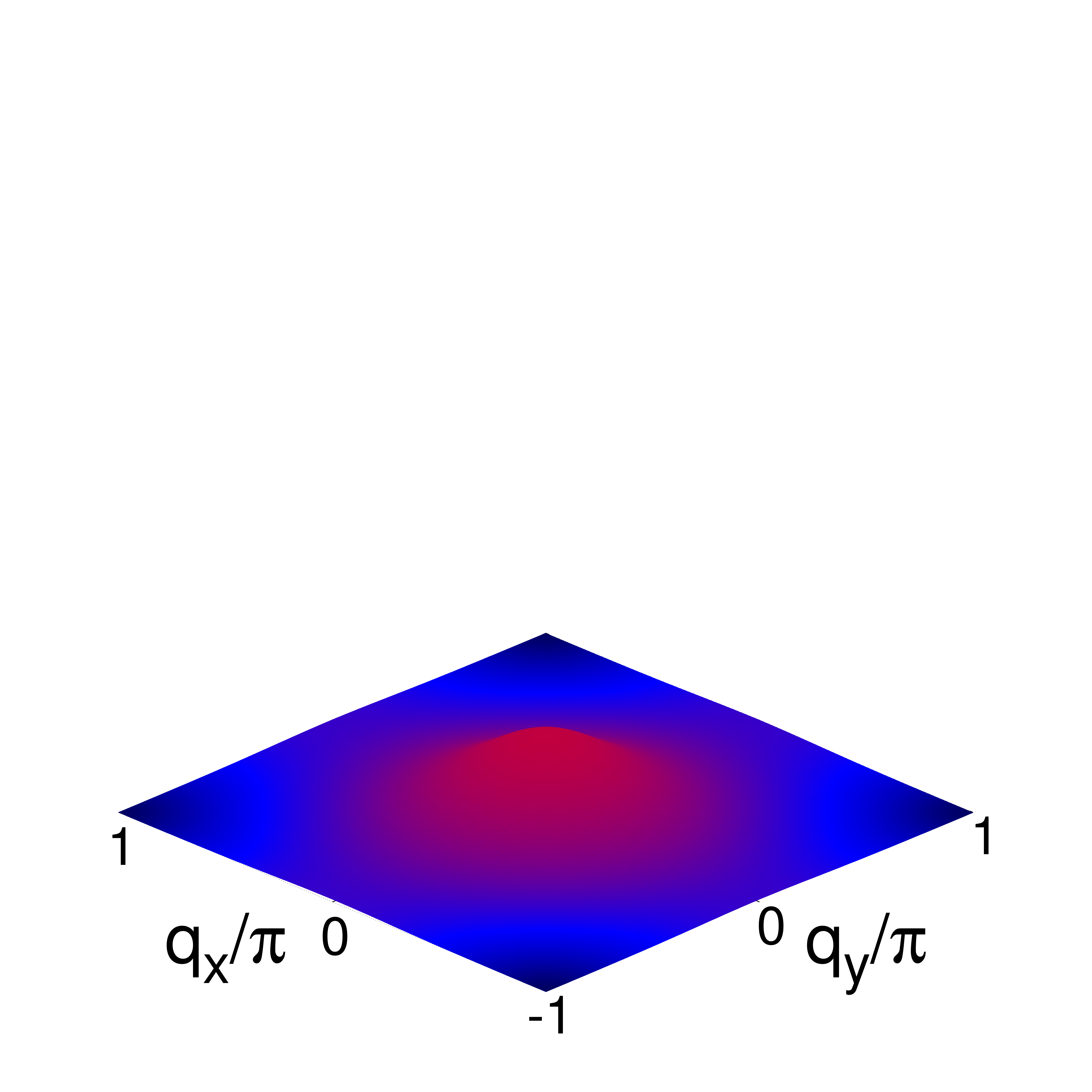}
\end{minipage}
\caption{(Upper panels) Momentum-resolved exciton densities $N_X(\q)$ and (lower panels) the corresponding momentum-resolved densities of correlated pairs $D_X(\q)$ for $T=0.4$ (cf. with marks in Fig.~\ref{PD_bandstructure}). The $z$-axis is equal within a vertical subfigure, but differs between the subfigures.}
\label{X_corr}
\end{figure}
Excitons can produce correlated pairs with the same momentum only, see Fig.~\ref{X_corr}. The ratio $D_X(\q)/N_X(\q)$ modulates with $\q$, suggesting that the size of the excitons depends on their momentum.

To analyze which electrons become correlated in the course of exciton formation, we resolve  $n_\downarrow^{\rm corr}$ according to the momenta of the electrons, $n_\downarrow^{\rm corr}=N^{-1}\sum_\k n_{\k\downarrow}^{\rm corr}$, with
\begin{equation}
n_{\k\downarrow}^{\rm corr} = \frac{U^2}{N} \sum_\q Z(\omega_X,\q) \frac{ p(\omega_X)
\big[f(\teps_{\k-\q\uparrow})-f(\teps_{\k\downarrow})\big] + f(\teps_{\k\downarrow})
\big[f(\teps_{\k-\q\uparrow})- 1 \big]} 
{ (\omega_X+\teps_{\k-\q\uparrow}-\teps_{\k\downarrow})^2}  .
\label{nCorr_k}
\end{equation}
The absolute values of $n_{\k\downarrow}^{\rm corr}$ can differ substantially. To compare the situations on the SM side with that on the SC side, we scale $n_{\k\downarrow}^{\rm corr}$ with its largest contribution, $\bar n_{\k\downarrow}^{\rm corr}=n_{\k\downarrow}^{\rm corr}/{\rm max}_\k^\pdagger  (n_{\k\downarrow}^{\rm corr})$.

\begin{figure}[h]
\centering
\subfigure{\includegraphics[height=0.25\linewidth]{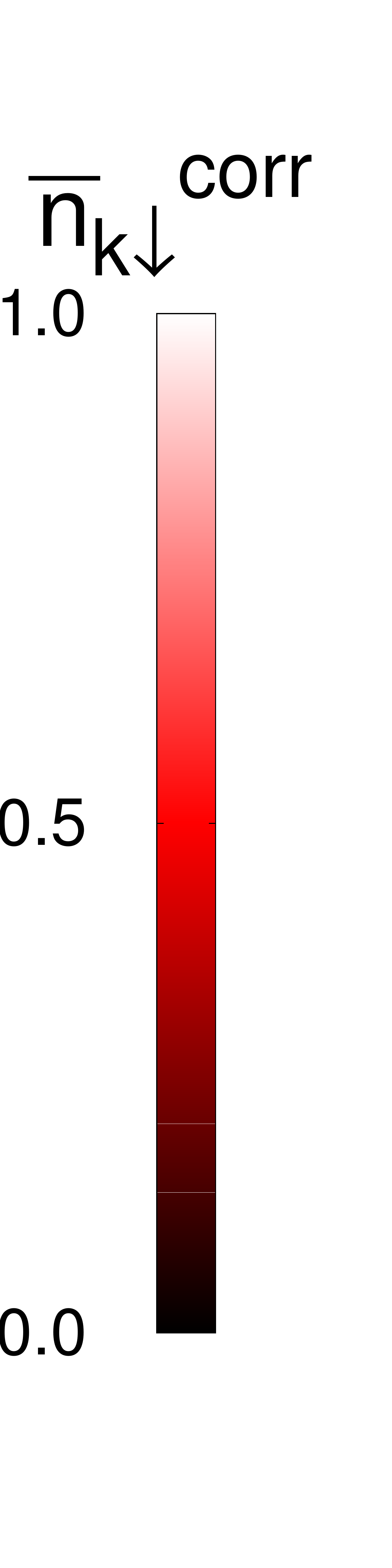}}
\setcounter{subfigure}{0}
\subfigure{\includegraphics[height=0.25\linewidth]{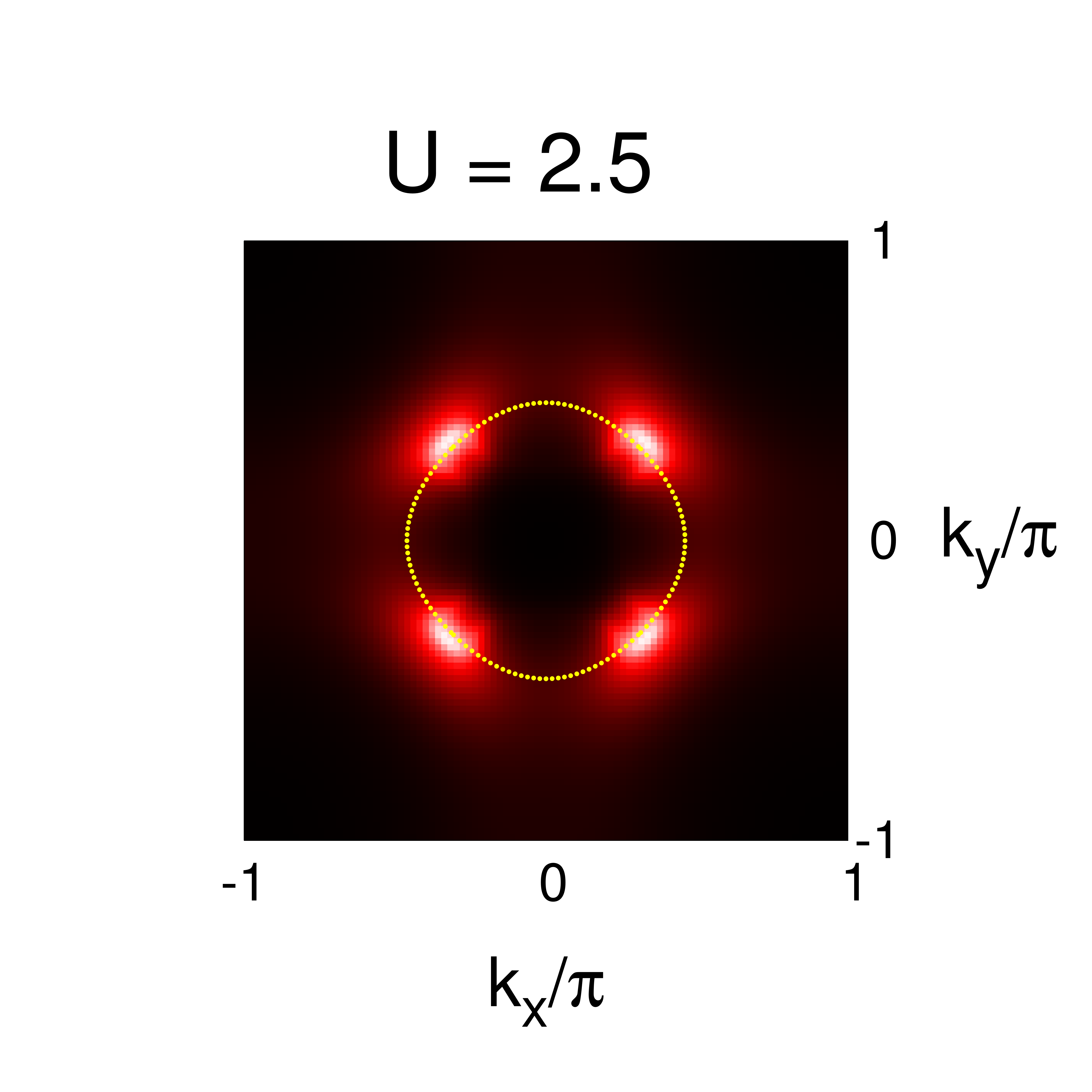}}
\subfigure{\includegraphics[height=0.25\linewidth]{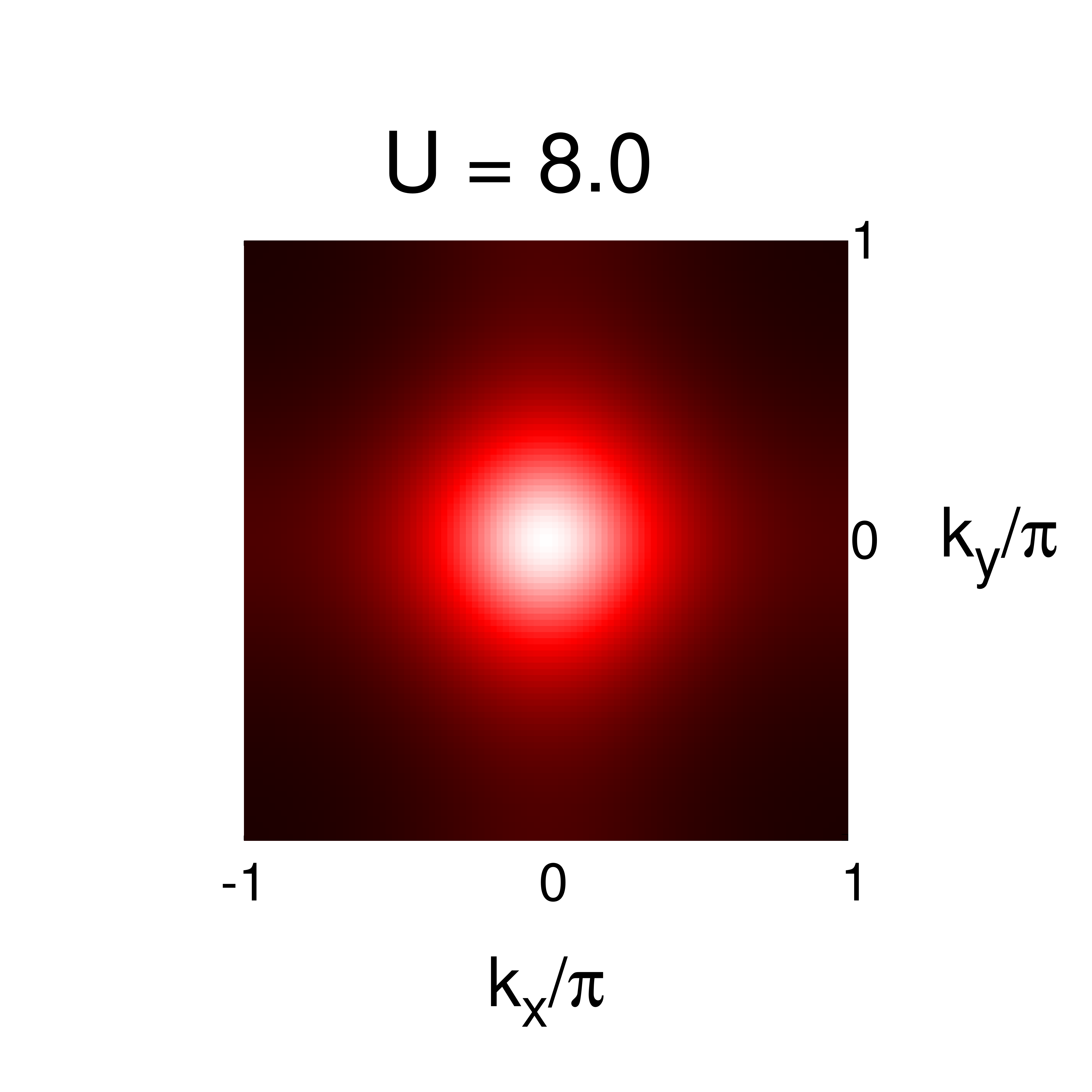}}
\caption{Momentum-resolved density of correlated electrons in case of (left panel) a  semimetal and (right panel) a semiconductor for $T=0.4$. The yellow, dotted line in the left panel shows the Fermi surface.}
\label{fig:nCorr_k}
\end{figure}

On the SM side, $\bar n_{\k\downarrow}^{\rm corr}$ shows a complicated structure. Let us recall that in a SM only finite-momentum excitons and correlated pairs exist, see Fig.~\ref{X_corr}(a). Here, the nesting property for finite momenta is crucial. Electrons become correlated only if holes with an adequate momentum are available. This is only the case at certain momenta close to the Fermi surface, see Fig.~\ref{fig:nCorr_k} ($U=2.5$). Moreover, $\bar n_{\k\downarrow}^{\rm corr}$ is sharply peaked, that is, the mean distance between the correlated electrons is large.

On the SC side, the zero-momentum correlations are most relevant (see $U=8.0$ in Fig.~\ref{X_corr}) and electrons close to the conduction-band minimum are predominantly correlated, as can be seen in Fig.~\ref{fig:nCorr_k} ($U=8.0$). The structure of $\bar n_{\k\downarrow}^{\rm corr}$ is broader than on the SM side, suggesting a more local nature of the correlations.
This corresponds to the situation within the EI phase, see the coherence length in Ref.~\cite{ZIBF12}.

\section{Conclusions}
\label{sec:Summary}
In this work we studied the normal phase close to the excitonic insulator (EI) phase, the so-called ``halo" phase. We focused on the correlations that are induced by the presence of excitons and calculated the self-energy by the projection technique for Green functions. This enabled us to separate the electron densities into a nearly-free part, which is in principle uncorrelated (only the band dispersion is renormalized), and a correlation part, where the influence of the excitons enters.

On the semimetal (SM) side excitons are loosely bound objects, whose radius is relatively large. Here, bound electron-hole pairs (excitons) only exist with finite momentum, and correlated electrons and holes form due to the presence of these excitons. The SM-EI transition is driven by critical electron-hole fluctuations. Due to the nesting property of the Fermi surface for finite momenta, correlated electrons are concentrated at four distinguished points close to the Fermi momenta.

With increasing Coulomb interaction the transition from the SM to the semiconductor (SC) is crossed. Thereby, the band structure and consequently the electron-hole pair spectrum changes dramatically.

In a SC no Fermi surface exists, and the electrons close to the conduction-band minimum are involved in the electron-hole correlations.
The excitons on the SC side are tighter bound, which suggests that their radius is smaller than on the SM side. Besides those with finite momentum, zero-momentum electron-hole pairs  exist, which represent the largest contribution to the electron-hole pair spectrum. The BEC of these excitons drive the SC-EI transition.

The qualitative difference in the electron-hole correlations on the SM side and on the SC side anticipates the BCS-BEC crossover within the EI phase.

Let us finally stress that the present work  exclusively addresses the formation of excitons and the question how they affect unbound charge carriers. The exciton-exciton interaction and particularly the formation of biexcitons is another interesting subject, which deserves further study in the framework of our approach. The formation of excitonic molecules could be relevant near the SM-SC transition, where a collective biexcitonic phase might possibly  compete with the EI~\cite{MRON13}.

\ack
This work was supported by the Deutsche Forschungsgemeinschaft through SFB 652, project B5.

\section*{References}

\begin{thebibliography}{10}
\expandafter\ifx\csname url\endcsname\relax
  \def\url#1{{\tt #1}}\fi
\expandafter\ifx\csname urlprefix\endcsname\relax\def\urlprefix{URL }\fi
\providecommand{\eprint}[2][]{\url{#2}}

\bibitem{Mo61}
Mott N~F 1961 {\em Philos. Mag.\/} {\bf 6} 287

\bibitem{Kno63}
Knox R 1963 {\em Solid State Physics\/} ed Seitz F and Turnbull D (New York:
  Academic Press) p Suppl. 5 p. 100

\bibitem{KK65}
Keldysh L~V and Kopaev H~Y~V 1965 {\em Sov. Phys. Sol. State\/} {\bf 6} 2219

\bibitem{JRK67}
J\'{e}rome D, Rice T~M and Kohn W 1967 {\em Physical Review\/} {\bf 158} 462

\bibitem{HR68}
Halperin B~I and Rice T~M 1968 {\em Rev. Mod. Phys.\/} {\bf 40} 755

\bibitem{BSW91}
Bucher B, Steiner P and Wachter P 1991 {\em Phys. Rev. Lett.\/} {\bf 67} 2717

\bibitem{CMCBDGBAPBF07}
Cercellier H, Monney C, Clerc F, Battaglia C, Despont L, Garnier M~G, Beck H,
  Aebi P, Patthey L, Berger H and Forr\'{o} L 2007 {\em Phys. Rev. Lett.\/}
  {\bf 99} 146403

\bibitem{WSTMANTKNT09}
Wakisaka Y, Sudayama T, Takubo K, Mizokawa T, Arita M, Namatame H, Taniguchi M,
  Katayama N, Nohara M and Takagi H 2009 {\em Phys. Rev. Lett.\/} {\bf 103}
  026402

\bibitem{PNH13}
Perali A, Neilson D and Hamilton A~R 2013 {\em Phys. Rev. Lett.\/} {\bf 110}
  146803

\bibitem{LEKMSS04}
Littlewood P~B, Eastham P~R, Keeling J~M~J, Marchetti F~M, Simons B~D and
  Szymanska M~H 2004 {\em J. Phys. Condens. Matter\/} {\bf 16} S3597

\bibitem{BF06}
Bronold F~X and Fehske H 2006 {\em Phys. Rev. \rm{B}\/} {\bf 74} 165107

\bibitem{IPBBF08}
Ihle D, Pfafferott M, Burovski E, Bronold F~X and Fehske H 2008 {\em Phys. Rev.
  \rm{B}\/} {\bf 78} 193103

\bibitem{ZIBF12}
Zenker B, Ihle D, Bronold F~X and Fehske H 2012 {\em Phys. Rev. \rm{B}\/} {\bf 85}
  121102(R)

\bibitem{ZIBF11}
Zenker B, Ihle D, Bronold F~X and Fehske H 2011 {\em Phys. Rev. \rm{B}\/} {\bf 83}
  235123

\bibitem{PFB11}
Phan N~V, Fehske H and Becker K~W 2011 {\em Europhys. Lett.\/} {\bf 95} 17006

\bibitem{Ba02b}
Batista C~D 2002 {\em Phys. Rev. Lett.\/} {\bf 89} 166403

\bibitem{BGBL04}
Batista C~D, Gubernatis J~E, Bon\v{c}a J and Lin H~Q 2004 {\em Phys. Rev.
  Lett.\/} {\bf 92} 187601

\bibitem{SC08}
Schneider C and Czycholl G 2008 {\em Eur. Phys. J. \rm{B}\/} {\bf 64} 43

\bibitem{ZIBF10}
Zenker B, Ihle D, Bronold F~X and Fehske H 2010 {\em Phys. Rev. \rm{B}\/} {\bf 81}
  115122

\bibitem{PBF10}
Phan N~V, Becker K~W and Fehske H 2010 {\em Phys. Rev. \rm{B}\/} {\bf 81} 205117

\bibitem{SEO11}
Seki K, Eder R and Ohta Y 2011 {\em Phys. Rev. \rm{B}\/} {\bf 84} 245106

\bibitem{Fa08}
Farka\v{s}ovsk\'{y} P 2008 {\em Phys. Rev. \rm{B}\/} {\bf 77} 155130

\bibitem{Pl11}
Plakida N~M 2011 {\em Springer Series in Solid-State Sciences\/} {\bf 171} 173

\bibitem{MS59}
Martin P~C and Schwinger J 1959 {\em Phys. Rev.\/} {\bf 115} 1342

\bibitem{KB62}
Kadanoff L~P and Baym G 1962 {\em Quantum Statistical Mechanics\/} (Reading,
  Massachusetts: Benjamin/Cumming Publishing Company)

\bibitem{KKER86}
Kraeft W~D, Kremp D, Ebeling W and R\"{o}pke G 1986 {\em Quantum Statistics of
  Charged Particle Systems\/} (Akademie-Verlag Berlin)

\bibitem{MRON13}
Maezono R, R\'{i}os P~L, Ogawa T and Needs R~J 2013 {\em Phys. Rev. Lett.\/}
  {\bf 110} 216407

\end{thebibliography}
\bibliographystyle{iopart-num}
\providecommand{\newblock}{}

\end{document}